\documentclass[11pt, onecolumn, draftcls]{IEEEtran}
\usepackage{graphics, epsfig, psfrag, amsmath, amsfonts, amssymb}
\newtheorem{thm}{Theorem}

\newtheorem{prop}{Proposition}
\newtheorem{defn}{Definition}

\newtheorem{example}{Example}

\DeclareMathAlphabet{\eurm}{U}{eur}{m}{n}
\DeclareMathAlphabet{\mathbsf}{OT1}{cmss}{bx}{n}
\DeclareMathAlphabet{\mathssf}{OT1}{cmss}{m}{sl}
\DeclareMathAlphabet{\mathcsf}{OT1}{cmss}{sbc}{n}

\newcommand{\samplevalue}[1]{\eurm{\lowercase{#1}}}
\newcommand{\randomvalue}[1]{\eurm{\uppercase{#1}}}

\DeclareSymbolFont{bsfletters}{OT1}{cmss}{bx}{n}  
\DeclareSymbolFont{ssfletters}{OT1}{cmss}{m}{n}
\DeclareMathSymbol{\bsfGamma}{0}{bsfletters}{'000}
\DeclareMathSymbol{\ssfGamma}{0}{ssfletters}{'000}
\DeclareMathSymbol{\bsfDelta}{0}{bsfletters}{'001}
\DeclareMathSymbol{\ssfDelta}{0}{ssfletters}{'001}
\DeclareMathSymbol{\bsfTheta}{0}{bsfletters}{'002}
\DeclareMathSymbol{\ssfTheta}{0}{ssfletters}{'002}
\DeclareMathSymbol{\bsfLambda}{0}{bsfletters}{'003}
\DeclareMathSymbol{\ssfLambda}{0}{ssfletters}{'003}
\DeclareMathSymbol{\bsfXi}{0}{bsfletters}{'004}
\DeclareMathSymbol{\ssfXi}{0}{ssfletters}{'004}
\DeclareMathSymbol{\bsfPi}{0}{bsfletters}{'005}
\DeclareMathSymbol{\ssfPi}{0}{ssfletters}{'005}
\DeclareMathSymbol{\bsfSigma}{0}{bsfletters}{'006}
\DeclareMathSymbol{\ssfSigma}{0}{ssfletters}{'006}
\DeclareMathSymbol{\bsfUpsilon}{0}{bsfletters}{'007}
\DeclareMathSymbol{\ssfUpsilon}{0}{ssfletters}{'007}
\DeclareMathSymbol{\bsfPhi}{0}{bsfletters}{'010}
\DeclareMathSymbol{\ssfPhi}{0}{ssfletters}{'010}
\DeclareMathSymbol{\bsfPsi}{0}{bsfletters}{'011}
\DeclareMathSymbol{\ssfPsi}{0}{ssfletters}{'011}
\DeclareMathSymbol{\bsfOmega}{0}{bsfletters}{'012}
\DeclareMathSymbol{\ssfOmega}{0}{ssfletters}{'012}

\newcommand{\rvX}{{\randomvalue{X}}}  	
\newcommand{\rvY}{{\randomvalue{Y}}}	
\newcommand{\rvZ}{{\randomvalue{Z}}}	

\newcommand{\rvh}{{\randomvalue{h}}}	

\newcommand{\rvt}{{\randomvalue{t}}}	

\newcommand{\rvx}{{\randomvalue{x}}}	
\newcommand{\svx}{{\samplevalue{x}}}

\newcommand{\rvy}{{\randomvalue{y}}}	
\newcommand{\svy}{{\samplevalue{y}}}

\newcommand{\rvz}{{\randomvalue{z}}}	

\newcommand{\svz}{{\samplevalue{z}}}

\begin{document}

\title{Orthogonal Codes for Robust Low-Cost Communication}
\author{\normalsize{Wenyi Zhang, {\it Member, IEEE}, and Urbashi Mitra, {\it Fellow, IEEE}}
\thanks{This work was supported in part by NSF through grants NRT ANI-0335302, ITR CCF-0313392, and OCE-0520324. The material in this paper was presented in part at the Information Theory and Applications (ITA) Workshop, San Diego, CA, January 2008.}
\thanks{W. Zhang was with the Ming Hsieh Department of Electrical Engineering, University of Southern California, Los Angeles, CA 90089, USA; and he is now with the Qualcomm Research Center, 5775 Morehouse Drive, San Diego, CA 92121, USA. Email: {\tt wenyizha@gmail.com}}
\thanks{U. Mitra is with the Ming Hsieh Department of Electrical Engineering, University of Southern California, Los Angeles, CA 90089, USA. Email: {\tt ubli@usc.edu}}
}

\maketitle

\begin{abstract}
Orthogonal coding schemes, known to asymptotically achieve the capacity per unit cost (CPUC) for single-user ergodic memoryless channels with a zero-cost input symbol, are investigated for single-user compound memoryless channels, which exhibit uncertainties in their input-output statistical relationships. A minimax formulation is adopted to attain robustness. First, a class of achievable rates per unit cost (ARPUC) is derived, and its utility is demonstrated through several representative case studies. Second, when the uncertainty set of channel transition statistics satisfies a convexity property, optimization is performed over the class of ARPUC through utilizing results of minimax robustness. The resulting CPUC lower bound indicates the ultimate performance of the orthogonal coding scheme, and coincides with the CPUC under certain restrictive conditions. Finally, still under the convexity property, it is shown that the CPUC can generally be achieved, through utilizing a so-called mixed strategy in which an orthogonal code contains an appropriate composition of different nonzero-cost input symbols.
\end{abstract}
\begin{keywords}
Capacity per unit cost, channel uncertainty, compound channel, minimax robustness, minimax theorem, mixed strategy, orthogonal code
\end{keywords}

\newpage

\section*{Abbreviations, Notations and Symbols}

\noindent \underline{ARPUC}: Achievable rate per unit cost, denoted $\mathbf{R}_\cdot$\\
\underline{CPUC}: Capacity per unit cost, denoted $\mathbf{C}_\cdot$\\
\underline{KL distance}: Kullback-Leibler distance, denoted $D(P\|Q)$ for two distributions $P$ and $Q$\\
$A := B$: Expression $A$ is defined by expression $B$\\
$\cdot^\mathrm{T}$: Transpose of vector or matrix\\
$\cdot^\dag$: Conjugate transpose of vector of matrix\\
$\mathbf{E}_P[\cdot]$: Expectation with respect to distribution $P$\\
$\mathrm{tr}[\cdot]$: Trace of matrix\\
$\det(\cdot)$: Determinant of matrix\\
$\mathcal{N}(\cdot, \cdot)$: Gaussian distribution\\
$\mathcal{CN}(\cdot, \cdot)$: Circularly symmetric complex Gaussian distribution\\
Random variables: $\rvx, \rvy \ldots$\\
Random vectors/matrices: $\underline{\rvx}, \underline{\rvy} \ldots$\\
Realizations of random variables: $\svx, \svy \ldots$\\
Realizations of random vectors/matrices: $\underline{\svx}, \underline{\svy} \ldots$\\
Sets, or alphabets of random variables: $\mathcal{X}, \mathcal{Y}, \mathcal{S} \ldots$

\noindent All logarithms are to the natural base $e$, and information units are measured in nats.

\section{Introduction}
\label{sec:intro}

The channel capacity quantifies the maximum amount of information transferable over a channel, measured on the basis of unit channel use (for discrete-time channels) or unit time duration (for continuous-time channels). Similarly, the channel capacity per unit cost (CPUC), initially investigated in a systematic way in \cite{verdu90:it},\footnote{The concept of CPUC was also explicitly proposed earlier in \cite[Chap. 2, Sec. 1, Ex. 26]{csiszar81:book}.} quantifies the maximum amount of information transferable over a channel, measured on the basis of unit average input cost. The relevance of the CPUC to communication systems stems from the fact that it directly quantifies the minimum cost necessary for reliably transmitting/receiving a unit of information, {\it e.g.}, the minimum energy per bit \cite{verdu02:it}.

In \cite{verdu90:it}, the CPUC of ergodic memoryless channels is established. Beyond a general single-letter characterization of the CPUC, it is further shown that when a zero-cost (a.k.a.~``free'') symbol is present in the input alphabet, the CPUC yields a particularly simple form involving the maximization of the Kullback-Leibler (KL) distance between two conditional output distributions.

In this paper, we consider the CPUC problem for channels with uncertain transition statistics. In particular, we adopt a single-user compound memoryless channel model. The compound channel model, in which the actual realization of the channel transition distributions is arbitrarily chosen from a parametrized uncertainty set and is not revealed to either the transmitter or the receiver, was initially introduced in \cite{blackwell59:ams} \cite{dobrushin59:ree}. The channel capacity was established therein,\footnote{The strong converse of the capacity theorem was established in \cite{wolfowitz60:arma}. Herein we will not elaborate upon the strong converse.} which takes a minimax form and can be interpreted as the outcome of a game between the input distribution and the channel realization.

By generalizing the proof of the CPUC formula for ergodic memoryless channels in a relatively straightforward manner, we obtain the CPUC general formula for compound memoryless channels with discrete alphabets. We subsequently turn to the important case where a unique zero-cost symbol is present in the input alphabet, and in particular investigate the achievable rates per unit cost (ARPUC) of an orthogonal coding scheme. While it is possible to use a sufficiently long block of training symbols to facilitate the receiver to achieve an essentially perfect knowledge of the channel transition statistics, throughout the paper, we make the assumption that there is no such ``training'' phase, and thus the receiver does not expend any effort in identifying the channel. From a practical perspective, it is often convenient to build communication systems that operate assuming some nominal channel model, or even simply ignore the actual channel statistics at all, due to the fact that precise identification of the underlying channel may demand excess resources and may be infeasible in certain situations. The orthogonal coding scheme also differs from other (possibly more powerful) coding schemes, like those using the maximum empirical mutual information (MMI) decoder \cite{csiszar81:book}, which not only achieves the capacity of compound memoryless channels, but also achieves their maximum random-coding error exponents. However, such ``universal'' coding schemes may be challenging to implement due to excessive decoder complexity for current communication systems (see, {\it e.g.}, \cite{lapidoth98:it} and references therein for a comprehensive overview of information-theoretic aspects of channels with various types of uncertainty).

Our investigation of the behavior of orthogonal codes is also motivated by the following considerations. First, since orthogonal codes asymptotically achieve the CPUC for ergodic memoryless channels with a zero-cost input symbol \cite{verdu90:it}, it is natural to ask whether a similar or disparate result holds for compound channels. Second, since orthogonal codes coupled with energy detection are CPUC-achieving simultaneously for a wide class of wideband fading Gaussian channels \cite{pierce66:it} (see also \cite{telatar00:it}), it is desirable to examine the validity of such a property for more general channels.

Part of the work of \cite{ganti00:it} is loosely related to our work. Besides other results, \cite{ganti00:it} investigates the ARPUC of mismatched decoding with additive decoding metrics \cite{csiszar95:it}. Specifically, the coding scheme considered therein is non-orthogonal, and the decoding algorithm depends on the sum of channel metrics that are independent of symbol positions in codewords. Thus it is unclear how to compare the ARPUC in \cite{ganti00:it} and that obtained in our work -- this is an interesting topic for future investigation.

We briefly summarize the content of the paper. First, we introduce the CPUC problem for compound memoryless channels, and for discrete alphabets present a general CPUC formula, which is a slight generalization of the corresponding result for ergodic memoryless channels. Subsequently, the focus is turned to orthogonal codes for channels with a zero-cost input symbol. By modifying the decoding algorithm from fixed-thresholding (from Stein's lemma) \cite{verdu90:it} to maximum-seeking, we derive an ARPUC formula, which depends upon the choice of processing function of channel outputs, and upon the choice of nonzero-cost input symbol.

We illustrate the utility of the derived ARPUC formula through several representative case studies. The aim is to show that the derived ARPUC formula provides a unified approach to evaluating and optimizing performance of cost-efficient communication systems, with different channel models and different receiver structures. Some additional insights regarding the behavior of these exemplar systems are also obtained by this exercise. We show that under certain conditions, simple receivers indeed yield rather good performance for corresponding compound memoryless channels. Specifically, we have the following observations. (a) Linear receivers achieve the CPUC for certain vector Gaussian additive-noise channels with partially unknown noise covariance, and guarantee the Gaussian-channel performance for scalar non-Gaussian noise channels. (b) Quadratic receivers achieve the CPUC for certain multipath channels with partially unknown multipath profile, and guarantee a Rayleigh-fading performance for non-coherent fading Gaussian channels with only the fading covariance matrix known. (c) Photon-counting receivers achieve the CPUC for certain Poisson channels with uncertainty in the background photon flow rate. Meanwhile, besides the preceding positive results, we also examine the lack of robustness of those receivers under certain other circumstances.

In order to systematically investigate the ultimate performance of orthogonal codes, we derive the maximum ARPUC by optimizing the processing function of channel outputs as well as the nonzero-cost input symbol. The maximization is analytically tractable when the uncertainty set of channel transition statistics satisfies a convexity property, with the aid of results from minimax robustness \cite{verdu84:it}. Similar to the CPUC for ergodic memoryless channels, the maximized ARPUC also involves the KL distance between two conditional output distributions. Furthermore, here the optimization requires solving a minimax game, which may be interpreted as one where the communication party initially selects a nonzero-cost input symbol in the orthogonal code, and then nature responds by selecting the least favorable channel state. The maximized ARPUC thus coincides with the CPUC if the aforementioned game has an equilibrium, but may also be strictly lower than the CPUC. Furthermore, analogous to the fact that mixed strategies enforce the existence of Nash equilibriums in finite games \cite{nash50:pnas}, in our problem, by adopting a mixed strategy in which an orthogonal code contains an appropriate composition of different nonzero-cost input symbols, we can further improve the ARPUC and indeed achieve the CPUC under the convexity condition.

The remainder of the paper is organized as follows. In Section \ref{sec:model} we introduce the CPUC problem, and present a general CPUC formula. In Section \ref{sec:orthogonal} we describe the orthogonal coding scheme, and derive an ARPUC formula. In Section \ref{sec:application} we illustrate the application of the derived ARPUC formula through several case studies. In Section \ref{sec:capacity} we perform optimization for the derived ARPUC formula, and obtain the corresponding CPUC lower bound and the CPUC that are achievable by orthogonal codes without and with the mixed strategy, respectively. In Section \ref{sec:conclusion} we conclude the paper.

\section{Channel Model and General CPUC Formula}
\label{sec:model}

We consider a single-user discrete-time compound memoryless channel with input $\rvx$ and output $\rvy$, with alphabets $\mathcal{X}$ and $\mathcal{Y}$, respectively. The channel has a parameter called the channel state $s \in \mathcal{S}$. As the channel state realization is fixed as $s$ throughout the entire coding block, the channel is memoryless and its input-output relationship is characterized by the conditional probability distribution $P_s(\svy|\svx)$ for input $\svx$ and output $\svy$, parametrized by $s$. As will be explicitly specified in the paper, the alphabets $\mathcal{X}$ and $\mathcal{Y}$ may be certain discrete and finite sets, or correspond to continuous sets such as the real line (then $P_s(\svy|\svx)$ should be understood as a probability density function). Every input symbol $\svx$ is associated with a cost function $c(\svx) \geq 0$. We adopt an additive cost structure such that for a block of channel inputs $(\svx_1, \svx_2, \ldots, \svx_n)$, the associated total cost is $\sum_{i = 1}^n c(\svx_i)$. Summarizing the preceding description, we specify the compound memoryless channel model as $\mathbsf{M}_\mathrm{c} = (\mathcal{X}, \mathcal{Y}, \mathcal{S}, P_{\cdot}(\cdot|\cdot), c(\cdot))$.

We note that the above channel model also encompasses a class of block interference channels \cite{mceliece84:it}, in which there are two states: -- a compound channel state that remains constant for all blocks, and a blockwise channel state that independently changes from block to block. This is because, conditioned upon a given compound channel state, the blockwise channel state may be absorbed in the channel transition distribution by treating each block as a supersymbol. For an example of such type of channels, consider a block fading channel \cite{marzetta99:it} with a partially unknown fading distribution, which corresponds to the compound channel state, and statistically governs the actual channel fading realizations corresponding to the blockwise channel state.

When the channel state set consists of a single element, $\mathcal{S} = \{s_0\}$, the compound memoryless channel reduces to an ergodic memoryless channel, and we write the channel model as $\mathbsf{M}_0 = (\mathcal{X}, \mathcal{Y}, P(\cdot|\cdot), c(\cdot))$ where the subscript of $P(\cdot|\cdot)$, $s_0$, is dropped.

We introduce the following definitions by adapting that for ergodic memoryless channels in \cite{verdu90:it}. For a given channel $\mathbsf{M}_\mathrm{c}$ (including $\mathbsf{M}_0$), an $(n, M, \nu, \epsilon)$ code is one in which the block length is equal to $n$; the number of messages is equal to $M$; corresponding to each message, the codeword $(\svx^{(m)}_1, \ldots, \svx^{(m)}_n)$, $m = 1, \ldots, M$, satisfies the constraint $\sum_{i = 1}^n c(\svx^{(m)}_i) \leq \nu$; and the average (over the ensemble of equiprobable messages) probability of decoding error is no greater than $\epsilon$.

\begin{defn}(\cite{blackwell59:ams})
Given $0 < \epsilon < 1$ and $\beta > 0$, a nonnegative number $R_\mathrm{c}$ is an $\epsilon$-achievable rate with cost per symbol not exceeding $\beta$ if for every $\gamma > 0$ there exists $n_0$ such that if $n \geq n_0$, then an $(n, M, n\beta, \epsilon)$ code can be found with $\log M > n(R_\mathrm{c} - \gamma)$, for every possible channel state $s \in \mathcal{S}$. If $R_\mathrm{c}$ is $\epsilon$-achievable for all $0 < \epsilon < 1$, it is an achievable rate; and the capacity under average cost $\beta$, $C_\mathrm{c}(\beta)$, is the maximum achievable rate.
\end{defn}

\begin{defn}
\label{def:cap-uncertainty}
Given $0 < \epsilon < 1$, a nonnegative number $\mathbf{R}_\mathrm{c}$ is an $\epsilon$-achievable rate per unit cost if for every $\gamma > 0$, there exists $\nu_0 \geq 0$ such that if $\nu \geq \nu_0$, then an $(n, M, \nu, \epsilon)$ code can be found with $\log M > \nu (\mathbf{R}_\mathrm{c} - \gamma)$, for every possible channel state $s \in \mathcal{S}$. If $\mathbf{R}_\mathrm{c}$ is $\epsilon$-achievable per unit cost for all $0 < \epsilon < 1$, it is an achievable rate per unit cost (ARPUC); and the capacity per unit cost (CPUC) $\mathbf{C}_\mathrm{c}$ is the maximum ARPUC.
\end{defn}

For ergodic memoryless channels, the following proposition establishes a general formula of the CPUC.
\begin{prop}(\cite[Theorem 2]{verdu90:it})
\label{prop:general-matched}
For an ergodic memoryless channel $\mathbsf{M}_0 = (\mathcal{X}, \mathcal{Y}, P(\cdot|\cdot), c(\cdot))$, the CPUC is
\begin{eqnarray}
\mathbf{C}_0 = \sup_{\stackrel{\beta > 0, P_\rvx}{\mathbf{E}[c(\rvx)] \leq \beta}} \frac{I(\rvx; \rvy)}{\beta}.
\end{eqnarray}
\end{prop}

In parallel to Proposition \ref{prop:general-matched}, the following proposition establishes a general formula of the CPUC of compound memoryless channels with discrete and finite alphabets.
\begin{prop}
\label{prop:general}
For a compound memoryless channel $\mathbsf{M}_\mathrm{c} = (\mathcal{X}, \mathcal{Y}, \mathcal{S}, P_{\cdot}(\cdot|\cdot), c(\cdot))$, in which $\mathcal{X}$ and $\mathcal{Y}$ are discrete and finite sets, the CPUC is
\begin{eqnarray}
\label{eqn:cpuc-general}
\mathbf{C}_\mathrm{c} = \sup_{\stackrel{\beta > 0, P_\rvx}{\mathbf{E}[c(\rvx)] \leq \beta}} \inf_{P_s(\cdot|\cdot): s\in \mathcal{S}} \frac{I(\rvx; \rvy)}{\beta}.
\end{eqnarray}
\end{prop}
The proof of Proposition \ref{prop:general} is in Appendix \ref{app:proof-general}. It essentially follows the proof of Proposition \ref{prop:general-matched}, with the classical result of compound channel capacity \cite{blackwell59:ams} appropriately utilized.

\section{Orthogonal Coding Scheme for Compound Memoryless Channels}
\label{sec:orthogonal}

In the sequel, we assume that there exists a unique symbol $\svx_\mathrm{f}$ with cost zero, {\it i.e.}, $c(\svx_\mathrm{f}) = 0$, in the input alphabet $\mathcal{X}$. For ergodic memoryless channels, the following proposition gives a particularly simple formula for the CPUC.
\begin{prop}(\cite[Theorem 3]{verdu90:it})
\label{prop:matched}
For an ergodic memoryless channel $\mathbsf{M}_0 = (\mathcal{X}, \mathcal{Y}, P(\cdot|\cdot), c(\cdot))$ with a unique zero-cost input symbol $\svx_\mathrm{f}$, the CPUC is
\begin{eqnarray}
\label{eqn:cap-matched-free}
\mathbf{C}_0 = \sup_{\svx \in \mathcal{X}^\prime} \frac{D\left(P(\svy|\svx)\| P(\svy|\svx_\mathrm{f})\right)}{c(\svx)},
\end{eqnarray}
where $\mathcal{X}^\prime := \mathcal{X} - \{\svx_\mathrm{f}\}$ and $D(\cdot\|\cdot)$ is the KL distance.
\end{prop}

In Section \ref{sec:capacity}, we will establish analogous results for compound memoryless channels that satisfy a certain convexity condition. In this section, we focus on the performance of specific orthogonal coding schemes. An important observation from \cite{verdu90:it} is that a deterministic orthogonal code can be explicitly constructed which asymptotically achieves the CPUC in Proposition \ref{prop:matched} as the coding block length grows large.

\begin{figure}[ht]
\centerline{\includegraphics[scale=0.7]{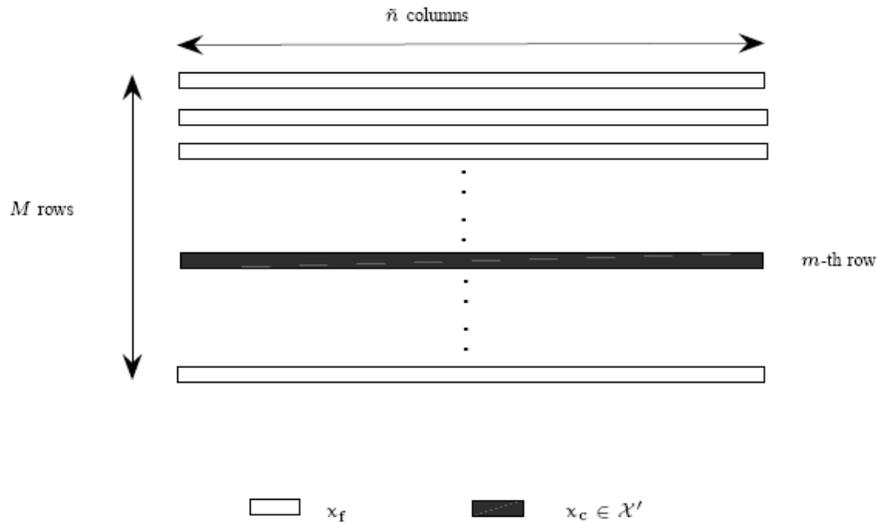}}
\caption{Illustration of an orthogonal codeword.}
\label{fig:orthocode}
\end{figure}

Except in Section \ref{subsec:mixed}, we will consider a simple orthogonal coding scheme without mixed strategy. The codebook construction and the corresponding decoding procedure are described as follows (see Figure \ref{fig:orthocode}).

{\it Codebook}:\\
Assume that the code consists of $M$ equally probable messages and that the coding block length is $n = M\tilde{n}$. Each codeword is virtually represented by an $M \times \tilde{n}$ two-dimensional block of channel uses. To represent message $m$, the $\tilde{n}$ elements in the $m$th row, $\{\rvX_{m, i}\}_{i = 1}^{\tilde{n}}$, take a symbol $\svx_\mathrm{c} \in \mathcal{X}^\prime$; and all the other $(M - 1)$ rows are all $\svx_\mathrm{f}$.

{\it Channel transmission}:\\
After the transmission of a codeword, the receiver observes $M$ mutually independent length-$\tilde{n}$ random vectors. Conditioned upon the compound channel state realization $s$, exactly one of the $M$ vectors consists of $\tilde{n}$ independent and identically distributed (i.i.d.) random variables following the distribution $P_{s, \mathrm{c}}(\svy) := P_s(\svy|\svx_\mathrm{c})$, and all the remaining following $P_{s, \mathrm{f}}(\svy) := P_s(\svy|\svx_\mathrm{f})$.

{\it Decoding}:\\
Given the received $M \times \tilde{n}$ block of channel outputs, if the receiver correctly finds the index of the row of inputs $\svx_\mathrm{c}$ we successfully decode the message, otherwise a decoding error occurs. For ergodic memoryless channels, a decoding algorithm based on Stein's lemma (see, {\it e.g.}, \cite[Chap.~12, Sec.~8]{cover91:book}) has been utilized in \cite{verdu90:it} to asymptotically achieve $\mathbf{C}_0$ as $\tilde{n} \rightarrow \infty$. The key component of that algorithm's implementation is a thresholding operation with threshold $\tau_\xi = D\left(P_\mathrm{c}(\svy)\|P_\mathrm{f}(\svy)\right) - \xi$ where $\xi > 0$ is made arbitrarily small as $\tilde{n} \rightarrow \infty$. For compound memoryless channels, such a threshold depends upon the channel state $s$, and thus the decoding algorithm is not applicable. Alternatively, we consider a decoding algorithm that computes a metric for each row of the received block and declares the decoded message as the row index with the maximum metric. Such an algorithm has been utilized in \cite{pierce66:it} for the particular case of fading Gaussian channels with energy detection, yielding an ARPUC identical to the CPUC of a Gaussian channel without fading, for any fading distributions with an identical second moment.

We now describe our decoding algorithm in detail. For computing the decoding metrics, the receiver first transforms each received channel output $\rvY_{m, i}$ into $g(\rvY_{m, i})$, $m = 1, \ldots, M$, $i = 1, \ldots, \tilde{n}$, where $g(\cdot)$ is an arbitrary real function that is measurable with respect to the probability measures $P_{s, \mathrm{f}}(\svy)$ and $P_{s, \mathrm{c}}(\svy)$, and satisfies $\mathbf{E}_{P_{s, \mathrm{c}}}[g(\rvY)] < \infty$, for all $s \in \mathcal{S}$. The receiver then computes the decoding metrics as
\begin{eqnarray}
\rvt_m := \frac{1}{\tilde{n}} \cdot \sum_{i = 1}^{\tilde{n}} g(\rvY_{m, i}), \quad m = 1, \ldots, M.
\end{eqnarray}
The decoding rule is to declare the decoded message $\hat{m}$ as that which maximizes the $M$ decoding metrics, {\it i.e.},
\begin{eqnarray}
\hat{m} = \mathrm{arg}\max_{m = 1, \ldots, M} \rvt_m.
\end{eqnarray}
If there exists a tie, then an error is declared.

For the described orthogonal coding scheme, its ARPUC is given by the following proposition.
\begin{prop}
\label{prop:rate}
For a compound memoryless channel $\mathbsf{M}_\mathrm{c} = \left(\mathcal{X}, \mathcal{Y}, \mathcal{S}, P_{\cdot}(\cdot|\cdot), c(\cdot)\right)$, if:
\begin{itemize}
\item[(a)] the input alphabet has a unique zero-cost symbol $\svx_\mathrm{f}$;
\item[(b)] an orthogonal code is used with the nonzero-cost input symbol $\svx_\mathrm{c}$ and with the transformation function $g(\cdot)$, such that condition
\begin{eqnarray}
\label{eqn:threshold-condition}
\inf_{s \in \mathcal{S}} \left\{\mathbf{E}_{P_{s, \mathrm{c}}} [g(\rvY)] - \mathbf{E}_{P_{s, \mathrm{f}}} [g(\rvY)]\right\} > 0
\end{eqnarray}
is satisfied,
\end{itemize}
then the ARPUC is
\begin{eqnarray}
\label{eqn:arpuc}
\mathbf{R}_\mathrm{c}(\svx_\mathrm{c}, g) = \inf_{s\in \mathcal{S}} \sup_{\theta \geq 0} \frac{1}{{c(\svx_\mathrm{c})}} \cdot\left\{
\theta \mathbf{E}_{P_{s, \mathrm{c}}} [g(\rvY)] - \log \mathbf{E}_{P_{s, \mathrm{f}}} \left[\exp(\theta g(\rvY))\right]
\right\}.
\end{eqnarray}
If condition (\ref{eqn:threshold-condition}) is violated, then the ARPUC is zero.
\end{prop}
{\it Proof}: The decoding error probability is equal to
\begin{eqnarray*}
\mathrm{Pr}\left[
\rvt_1 \leq \max_{m = 2, \ldots, M} \rvt_m
\right],
\end{eqnarray*}
where for $\rvt_1$, the channel outputs $\{\rvY_{1, i}\}_{i = 1}^{\tilde{n}}$ are distributed as $P_{s, \mathrm{c}}(\svy)$, and for $\rvt_{m \neq 1}$ the channel outputs $\{\rvY_{m, i}\}_{i = 1}^{\tilde{n}}$ are distributed as $P_{s, \mathrm{f}}(\svy)$. Now let us examine the conditions for the decoding error probability to vanish as $\tilde{n}$ goes large.

A prerequisite is that, the expectation of $\rvt_1$ should be strictly greater than that of $\rvt_{m \neq 1}$ for all possible channel state realizations, leading to
\begin{eqnarray}
\inf_{s \in \mathcal{S}} \left\{\mathbf{E}_{P_{s, \mathrm{c}}} [g(\rvY)] - \mathbf{E}_{P_{s, \mathrm{f}}} [g(\rvY)]\right\} > 0.
\end{eqnarray}

Then let us fix a channel state realization $s$. Since $\rvt_1$ is the empirical mean of $\tilde{n}$ i.i.d. random variables, by the weak law of large numbers,
\begin{eqnarray}
\lim_{\tilde{n} \rightarrow \infty} \mathrm{Pr}\left[
|\rvt_1 - \mathbf{E}_{P_{s, \mathrm{c}}}[g(\rvY)]| < \xi
\right] = 1
\end{eqnarray}
for any $\xi > 0$. So for arbitrarily small $\epsilon > 0$ and $\xi > 0$, there exists a positive integer $\tilde{n}_1$ such that for every $\tilde{n} \geq \tilde{n}_1$,
\begin{eqnarray}
\mathrm{Pr}\left[
\rvt_1 \leq \mathbf{E}_{P_{s, \mathrm{c}}}[g(\rvY)] - \xi
\right] < \frac{\epsilon}{2}.
\end{eqnarray}

For $\max_{m = 2, \ldots, M} T_m$, the cumulative distribution function is bounded by
\begin{eqnarray}
\mathrm{Pr}\left[\max_{m = 2, \ldots, M} \rvt_m \leq t\right] &=& \left\{1 - \mathrm{Pr}\left[\rvt_2 > t\right]\right\}^{M - 1}\nonumber\\
&>& 1 - M \cdot\mathrm{Pr}\left[\rvt_2 > t\right].
\end{eqnarray}
The tail probability $\mathrm{Pr}\left[\rvt_2 > t\right]$ can be further upper bounded by the Chernoff bound as
\begin{eqnarray}
\mathrm{Pr}\left[\rvt_2 > t\right] \leq \exp\left\{-\tilde{n}\cdot \sup_{\theta \geq 0}
\left[\theta t - \log \mathbf{E}_{P_{s, \mathrm{f}}}[\exp(\theta g(\rvY))]\right]
\right\}.
\end{eqnarray}
Hence we have
\begin{eqnarray}
&&\mathrm{Pr}\left[\max_{m = 2, \ldots, M} \rvt_m > \mathbf{E}_{P_{s, \mathrm{c}}}[g(\rvY)] - \xi \right]\nonumber\\ &=& 1 - \mathrm{Pr}\left[\max_{m = 2, \ldots, M} \rvt_m \leq \mathbf{E}_{P_{s, \mathrm{c}}}[g(\rvY)] - \xi \right]\nonumber\\
&<& M \cdot \mathrm{Pr}\left[\rvt_2 > \mathbf{E}_{P_{s, \mathrm{c}}}[g(\rvY)] - \xi\right]\nonumber\\
&\leq& M\cdot \exp\left\{-\tilde{n}\cdot \sup_{\theta \geq 0}
\left[\theta (\mathbf{E}_{P_{s, \mathrm{c}}}[g(\rvY)] - \xi) - \log \mathbf{E}_{P_{s, \mathrm{f}}}[\exp(\theta g(\rvY))]\right]
\right\},
\end{eqnarray}
which can be made no greater than $\epsilon/2$ for every $\tilde{n} \geq \tilde{n}_2$ where $\tilde{n}_2$ is a sufficiently large integer, if the growth of $M$ satisfies
\begin{eqnarray}
\label{eqn:growth-M}
\frac{\log M}{\tilde{n}} = \sup_{\theta \geq 0}
\left[\theta (\mathbf{E}_{P_{s, \mathrm{c}}}[g(\rvY)] - \xi) - \log \mathbf{E}_{P_{s, \mathrm{f}}}[\exp(\theta g(\rvY))]\right] - \xi.
\end{eqnarray}

To complete the proof, we let $\epsilon \rightarrow 0$ and $\xi \rightarrow 0$, and take infimum of the right hand side of (\ref{eqn:growth-M}) over all possible $s \in \mathcal{S}$. Then the ARPUC (\ref{eqn:arpuc}) is achievable as $\tilde{n} \rightarrow \infty$. {\bf Q.E.D.}

{\it Remarks:}
\begin{itemize}
\item For ergodic memoryless channels, by letting $g(\rvY) = \log[P_\mathrm{c}(\rvY)/P_\mathrm{f}(\rvY)]$ and $\theta = 1$, the ARPUC (\ref{eqn:arpuc}) becomes
\begin{eqnarray}
\mathbf{R}_0 = \frac{1}{c(\svx_\mathrm{c})}\cdot\mathbf{E}_{P_\mathrm{c}}\left[\log \frac{P_\mathrm{c}(\rvY)}{P_\mathrm{f}(\rvY)}\right] = \frac{D\left(
P_\mathrm{c}(\svy)\|P_\mathrm{f}(\svy)
\right)}{c(\svx_\mathrm{c})}.
\end{eqnarray}
By optimizing over all possible nonzero-cost input symbols $\svx_\mathrm{c} \in \mathcal{X}^\prime$, we revisit the achievability of $\mathbf{C}_0$ in Proposition \ref{prop:matched}. We note that to achieve $\mathbf{R}_0$, either thresholding (\cite{verdu90:it}) or maximum-seeking as described above suffices.

\item As can be seen from its proof, Proposition \ref{prop:rate} is not based upon Proposition \ref{prop:general}, and thus does not require the alphabets $\mathcal{X}, \mathcal{Y}$ be discrete or finite. In Section \ref{sec:application}, we indeed evaluate the ARPUC for several channels with continuous channel outputs.

\item Proposition \ref{prop:rate} only establishes the achievability of the ARPUC. In fact, by utilizing Cram\'{e}r's theorem (see, {\it e.g.}, \cite[Theorem 2.2.3]{dembo98:book}), we can show that $\mathbf{R}_\mathrm{c}(\svx_\mathrm{c}, g)$ is also the maximum ARPUC for the orthogonal coding scheme with given $\svx_\mathrm{c}$ and $g(\cdot)$, as shown in Appendix \ref{app:arpuc-max}.

\item In the orthogonal coding scheme, we can optimize over all possible $g(\cdot)$ and $\svx_\mathrm{c} \in \mathcal{X}^\prime$, to maximize $\mathbf{R}_\mathrm{c}(\svx_\mathrm{c}, g)$ in Proposition \ref{prop:rate}. We will systematically perform the optimization in Section \ref{sec:capacity} for channels whose uncertainty satisfies a convexity property.
\end{itemize}

\section{Evaluation of ARPUC for Several Applications}
\label{sec:application}

In this section, we illustrate the utility of Proposition \ref{prop:rate} through a series of exemplar applications. We consider several representative receivers, applied to several practically important channel models. Hence these case studies are by no means purely academic exercises, but shed light on the behavior of certain communication systems under channel uncertainties. These case studies include linear and quadratic receivers for linear additive-noise channels, quadratic receivers for non-coherent fading channels, and photon-counting receivers for Poisson channels. We observe that, under various situations, Proposition \ref{prop:rate} provides a unified approach to evaluating and optimizing performance of these classes of receiver-channel pairs.

\subsection{Linear and Quadratic Receivers for Linear Additive-Noise Channels}

Consider the discrete-time memoryless additive-noise channel
\begin{eqnarray}
\underline{\rvy} = \underline{H}^\mathrm{T}\underline{\rvx} + \underline{\rvz},
\end{eqnarray}
where all the quantities are real-valued. The input $\underline{\rvx}$ is $n_\mathrm{t}$-dimensional, the output $\underline{\rvy}$ and the additive noise $\underline{\rvz}$ are $n_\mathrm{r}$-dimensional, and the deterministic channel transfer matrix $\underline{H}$ is $n_\mathrm{t} \times n_\mathrm{r}$. For simplicity, we assume that $\underline{\rvz}$ is independent of $\underline{\rvx}$ and has mean zero. Such a channel model encompasses multi-dimensional Gaussian channels ({\it e.g.}, \cite{tsybakov65:pit}, \cite{telatar98:ett}), additive-noise channels with non-Gaussian noises ({\it e.g.}, \cite{aazhang87:com}), channels with intersymbol interference (ISI) or multipath ({\it e.g.}, \cite{proakis00:book}).

A linear receiver combines components of the channel output vector, and may be readily implemented by a linear filter. It possesses the following structure,
\begin{eqnarray}
g_\mathrm{l}(\underline{\rvy}) = \underline{w}_\mathrm{l}^\mathrm{T} \underline{\rvy},
\end{eqnarray}
where $\underline{w}_\mathrm{l}$ is a deterministic $n_\mathrm{r}$-dimensional combining vector. Alternatively, a general quadratic receiver extracts signal energy, but ignores the phase information. It possesses the following structure,
\begin{eqnarray}
g_\mathrm{q}(\underline{\rvy}) = \underline{\rvy}^\mathrm{T} \underline{W}_\mathrm{q} \underline{\rvY},
\end{eqnarray}
where $\underline{W}_\mathrm{q}$ is a deterministic positive semi-definite matrix.

For the linear additive-noise channel, we define the cost function as the energy of an input, so that
\begin{eqnarray}
c(\underline{\rvx}) := \sum_{i = 1}^{n_\mathrm{t}} |\rvx_i|^2 = \|\underline{\rvX}\|^2,
\end{eqnarray}
and the zero-cost symbol is $\underline{\svx}_\mathrm{f} = \underline{0}$. The channel has two states with uncertainty. First, the channel transfer matrix $\underline{H}$ is arbitrarily drawn from a set $\mathcal{H}$. Second, the distribution of the additive noise, $P_{\underline{\rvz}}$, is arbitrarily drawn from a set of distributions, $\mathcal{P}_{\underline{\rvz}}$. So the channel state is denoted by $s = (\underline{H}, P_{\underline{\rvz}})$.

Applying Proposition \ref{prop:rate}, we can evaluate the ARPUC for linear and quadratic receivers, as follows, noting that $\underline{\svx}_\mathrm{c}$ is the nonzero-cost input symbol (vector) in the orthogonal coding scheme.
\begin{eqnarray}
&&\mathrm{Linear\; receivers:}\nonumber\\
\label{eqn:rate-additive}
\mathbf{R}_\mathrm{c}(\underline{\svx}_\mathrm{c}, \underline{w}_\mathrm{l}) &=& \inf_{s \in \mathcal{S}} \sup_{\theta \geq 0} \frac{1}{\|\underline{\svx}_\mathrm{c}\|^2}\left\{
\theta (\underline{H} \underline{w}_\mathrm{l})^\mathrm{T} \underline{\svx}_\mathrm{c} - \log \mathbf{E}[\exp(\theta \underline{w}_\mathrm{l}^\mathrm{T} \underline{\rvZ})]
\right\},\\
\mathrm{if}\quad&& \inf_{s \in \mathcal{S}} (\underline{H} \underline{w}_\mathrm{l})^\mathrm{T} \underline{\svx}_\mathrm{c} > 0; \;\mathrm{and\;zero\;otherwise}.\nonumber\\
&&\mathrm{Quadratic\; receivers:}\nonumber\\
\label{eqn:rate-additive-quadratic}
\mathbf{R}_\mathrm{c}(\underline{\svx}_\mathrm{c}, \underline{W}_\mathrm{q}) &=& \inf_{s \in \mathcal{S}} \sup_{\theta \geq 0} \frac{1}{\|\underline{\svx}_\mathrm{c}\|^2}\left\{
\theta \underline{\svx}_\mathrm{c}^\mathrm{T} \underline{H} \underline{W}_\mathrm{q} \underline{H}^\mathrm{T} \underline{\svx}_\mathrm{c} + \theta \mathbf{E}[\underline{\rvZ}^\mathrm{T} \underline{W}_\mathrm{q}\underline{\rvZ}] - \log \mathbf{E}[\exp(\theta \underline{\rvZ}^\mathrm{T} \underline{W}_\mathrm{q}\underline{\rvZ})]
\right\},\\
\mathrm{if}\quad&& \inf_{s \in \mathcal{S}} \underline{\svx}_\mathrm{c}^\mathrm{T} \underline{H} \underline{W}_\mathrm{q} \underline{H}^\mathrm{T} \underline{\svx}_\mathrm{c} > 0; \;\mathrm{and\;zero\;otherwise}.\nonumber
\end{eqnarray}

For concreteness, we examine the following specific examples in more detail.

\begin{example}{\it (Linear Receivers for Gaussian Noise Channels with Partially Unknown Covariance)}
\label{example:gaussian-covariance}
Let $n_\mathrm{t} = n_\mathrm{r}$, $\mathcal{H} = \{\underline{I}\}$, and $\underline{\rvZ} \sim \mathcal{N}(\underline{0}, \underline{\Phi})$ with positive-semidefinite covariance matrix $\underline{\Phi} \in \mathcal{S}_{\underline{\Phi}}$. The capacity of such a channel has been addressed as a special case of compound linear Gaussian channels in \cite{root68:siam}. Here we focus on its ARPUC and CPUC, and as will be shown, under a specific convexity condition, the CPUC can be achieved by the orthogonal coding scheme with a linear receiver. Another related problem has been addressed in \cite{diggavi01:it} (also see references therein), where the focus is on finding the worst noise distribution under a given covariance structure. Here in contrast, we focus on the case where the noise is Gaussian, but its covariance matrix is not perfectly identified, due to practical issues such as limited channel training or time-variations.

We consider linear receivers. The ARPUC can be shown to be
\begin{eqnarray}
\label{eqn:rate-vecGaussian}
\mathbf{R}_\mathrm{c}(\underline{\svx}_\mathrm{c}, \underline{w}_\mathrm{l}) &=& \inf_{\underline{\Phi} \in \mathcal{S}_{\underline{\Phi}}} \frac{(\underline{w}_\mathrm{l}^\mathrm{T}\underline{\svx}_\mathrm{c})^2}{2 (\underline{w}_\mathrm{l}^\mathrm{T}\underline{\Phi}\underline{w}_\mathrm{l})\|\underline{\svx}_\mathrm{c}\|^2},
\end{eqnarray}
if $\underline{w}_\mathrm{l}^\mathrm{T} \underline{\svx}_\mathrm{c} > 0$; and zero otherwise.

From (\ref{eqn:rate-vecGaussian}), we can proceed further to optimize the ARPUC over all feasible $\underline{\svx}_\mathrm{c}$ and $\underline{w}_\mathrm{l}$. From the Cauchy-Schwartz inequality, we have
\begin{eqnarray}
\mathbf{R}_\mathrm{c}(\underline{\svx}_\mathrm{c}, \underline{w}_\mathrm{l}) \leq \inf_{\underline{\Phi} \in \mathcal{S}_{\underline{\Phi}}} \frac{\|\underline{w}_\mathrm{l}\|^2}{2 (\underline{w}_\mathrm{l}^\mathrm{T}\underline{\Phi}\underline{w}_\mathrm{l})},
\end{eqnarray}
where equality is achieved if and only if $\underline{w}_\mathrm{l}$ is proportional to $\underline{\svx}_\mathrm{c}$. Therefore, the optimal ARPUC $\mathbf{R}_\mathrm{c}$ is
\begin{eqnarray}
\label{eqn:gaussian-saddlepoint}
\mathbf{R}_\mathrm{c} = \sup_{\|\underline{w}_\mathrm{l}\| = 1} \inf_{\underline{\Phi} \in \mathcal{S}_{\underline{\Phi}}} \frac{1}{2\underline{w}_\mathrm{l}^\mathrm{T}\underline{\Phi}\underline{w}_\mathrm{l}}.
\end{eqnarray}
Assume that the uncertainty set $\mathcal{S}_{\underline{\Phi}}$ is compact and convex, and that $\max_{\underline{\Phi} \in \mathcal{S}_{\underline{\Phi}}} \lambda_{\min}(\underline{\Phi})$ is lower bounded away from zero, where $\lambda_{\min}(\underline{\Phi})$ denotes the minimum eigenvalue of $\underline{\Phi}$. We can utilize results of minimax robustness to find that (\ref{eqn:gaussian-saddlepoint}) has a saddle point, so that
\begin{eqnarray}
\label{eqn:gaussian-covariance}
\mathbf{R}_\mathrm{c} = \frac{1}{2 \max_{\underline{\Phi} \in \mathcal{S}_{\underline{\Phi}}} \lambda_{\min}(\underline{\Phi})}.
\end{eqnarray}
We show the derivation of (\ref{eqn:gaussian-covariance}) in Appendix \ref{app:gaussian-covariance}. We note that (\ref{eqn:gaussian-covariance}) coincides with the CPUC under the ideal assumption that the transmitter and the receiver both have perfect knowledge of the realization of $\underline{\Phi}$. Therefore in this example we have shown that, for compact and convex $\mathcal{S}_{\underline{\Phi}}$, linear receivers indeed achieve the CPUC of Gaussian noise channels with partially unknown covariance.
\end{example}

\begin{example}{\it (Quadratic Receivers for Multipath Channels)}

Let $n_\mathrm{t} = 1$, $\underline{\rvZ} \in \mathcal{N}(0, \underline{I})$, and $\underline{H} \in \mathcal{H}$ models channel multipath. In many sparse multipath channels, $\underline{H}$ may have a dimension of several tens or higher with mostly zero elements, {\it e.g.}, \cite{fevrier99:com}, \cite{molisch:report}, \cite{song05:asaj}. For those situations, linear receivers like the RAKE receiver ({\it e.g.}, \cite{proakis00:book}) can be sensitive to channel estimation errors. Alternatively, quadratic receivers may be employed to detect signal energy. By Proposition \ref{prop:rate}, for quadratic receivers the ARPUC can be shown as
\begin{eqnarray}
\label{eqn:mp-q}
\mathbf{R}_\mathrm{c}(\svx_\mathrm{c}, \underline{W}_\mathrm{q}) &=& \inf_{\underline{H} \in \mathcal{H}}\sup_{\theta\geq 0}\frac{1}{\svx_\mathrm{c}^2}\left\{
\theta \svx_\mathrm{c}^2 (\underline{H}\underline{W}_\mathrm{q}\underline{H}^\mathrm{T}) + \theta \mathrm{tr}[\underline{W}_\mathrm{q}] + \frac{1}{2}\log\det(\underline{I} - 2\theta \underline{W}_\mathrm{q}) \right\},
\end{eqnarray}
with the condition $\inf_{\underline{H}\in \mathcal{H}} (\underline{H}\underline{W}_\mathrm{q}\underline{H}^\mathrm{T}) > 0$ satisfied, where $\theta$ is bounded such that $\det(\underline{I} - 2\theta \underline{W}_\mathrm{q}) > 0$. If $\underline{W}_\mathrm{q} = \underline{I}$, (\ref{eqn:mp-q}) further becomes
\begin{eqnarray}
\label{eqn:mp-q-s}
\mathbf{R}_\mathrm{c}(\svx_\mathrm{c}, \underline{I}) = \inf_{\underline{H}\in \mathcal{H}}\left\{
\frac{\|\underline{H}\|^2}{2} - \frac{1}{2 \svx_\mathrm{c}^2/n_\mathrm{r}} \log\left(1 + \|\underline{H}\|^2 \svx_\mathrm{c}^2/n_\mathrm{r}
\right)
\right\}.
\end{eqnarray}
It is easily seen that (\ref{eqn:mp-q-s}) is maximized by letting $|\svx_\mathrm{c}| \rightarrow \infty$, as
\begin{eqnarray}
\mathbf{R}_\mathrm{c} = \inf_{\underline{H}\in \mathcal{H}} \frac{\|\underline{H}\|^2}{2},
\end{eqnarray}
which is also the CPUC under the ideal assumption that the transmitter and the receiver both have perfect knowledge of the realization of $\underline{H}$.

On the other hand, if there is a finite peak limit for $\svx_\mathrm{c}$, then as the channel dimension $n_\mathrm{r} \rightarrow \infty$, (\ref{eqn:mp-q-s}) becomes zero for any $\|\underline{H}\|^2 < \infty$. Intuitively, this is because as the channel dimension increases, the quadratic receiver tends to collect more noise than useful signals. In practice, a multipath channel often has distinct multipath delay profiles from realization to realization ({\it e.g.}, \cite{franz06:jsac}), so robust receivers designed here tend to yield rather conservative performance for most channel realizations, because they need to cope with channels with extremely long delay spreads, which occur with a rather small probability. A more plausible alternative may be an outage-based approach, which is beyond the scope of this paper.
\end{example}

\begin{example}{\it (Scalar Non-Gaussian Noise Channels)}
\label{example:non-Gaussian}

Let $n_\mathrm{t} = n_\mathrm{r} = 1$, $\mathcal{H} = \{1\}$, and the zero-mean additive noise $\rvZ$ have an unknown probability density function $P_\rvZ \in \mathcal{P}_\rvz$ with a common variance $\mathrm{var}(\rvZ) < \infty$. Without loss of generality, we can let $w_\mathrm{l} = W_\mathrm{q} = 1$, and the corresponding ARPUCs are
\begin{eqnarray}
\label{eqn:rate-nonG}
\mathbf{R}_\mathrm{c}(\svx_\mathrm{c}, 1) = \inf_{P_\rvZ} \sup_{\theta \geq 0} \frac{1}{\svx_\mathrm{c}^2}\left\{
\theta \svx_\mathrm{c} - \log \mathbf{E}\left[
\exp(\theta \rvZ)
\right]
\right\},\quad\mathrm{for\;linear\;receivers;}
\end{eqnarray}
and
\begin{eqnarray}
\label{eqn:rate-nonG-q}
\mathbf{R}_\mathrm{c}(\svx_\mathrm{c}, 1) = \inf_{P_\rvZ} \sup_{\theta \geq 0} \frac{1}{\svx_\mathrm{c}^2}\left\{
\theta (\svx_\mathrm{c}^2 + \mathrm{var}(\rvZ)) - \log \mathbf{E}[\exp(\theta \rvZ^2)]
\right\},\quad\mathrm{for\;quadratic\;receivers.}
\end{eqnarray}

For linear receivers, if we let $\svx_\mathrm{c} \rightarrow 0$ and $\theta = \svx_\mathrm{c}/\mathrm{var}(\rvZ)$ in (\ref{eqn:rate-nonG}), we get
\begin{eqnarray}
\label{eqn:nonG-guarantee}
\lim_{\svx_\mathrm{c}\rightarrow 0}\mathbf{R}_\mathrm{r}(\svx_\mathrm{c}, 1) &\geq& \lim_{\svx_\mathrm{c}\rightarrow 0}\inf_{P_\rvZ} \left\{
\frac{1}{\mathrm{var}(\rvZ)} - \frac{\log \mathbf{E}\left[\exp(\svx_\mathrm{c}\rvZ/\mathrm{var}(\rvZ))\right]}{\svx_\mathrm{c}^2}
\right\}\nonumber\\
&=& \frac{1}{\mathrm{var}(\rvZ)} - \lim_{\svx_\mathrm{c}\rightarrow 0} \frac{\sup_{P_\rvZ} \log \mathbf{E}\left[\exp(\svx_\mathrm{c}\rvZ/\mathrm{var}(\rvZ))\right]}{\svx_\mathrm{c}^2}\nonumber\\
&=& \frac{1}{\mathrm{var}(\rvZ)} - \lim_{\svx_\mathrm{c}\rightarrow 0} \frac{\sup_{P_\rvZ} \log \mathbf{E}\left[1 + \svx_\mathrm{c}\rvZ/\mathrm{var}(\rvZ) + (1/2)\cdot\svx_\mathrm{c}^2\rvZ^2/\mathrm{var}(\rvZ)^2 + o(\svx_\mathrm{c}^2)\right]}{\svx_\mathrm{c}^2}\nonumber\\
&=& \frac{1}{\mathrm{var}(\rvZ)} - \lim_{\svx_\mathrm{c}\rightarrow 0} \frac{\sup_{P_\rvZ} \log [1 + (1/2)\cdot\svx_\mathrm{c}^2/\mathrm{var}(\rvZ) + o(\svx_\mathrm{c}^2)]}{\svx_\mathrm{c}^2}\nonumber\\
&=& \frac{1}{\mathrm{var}(\rvZ)} - \frac{1}{2\cdot \mathrm{var}(\rvZ)} = \frac{1}{2\cdot\mathrm{var}(\rvZ)},
\end{eqnarray}
independent of the actual noise distribution, and identical to the CPUC of Gaussian channels whose noise variance is $\mathrm{var}(\rvz)$. The lower bound (\ref{eqn:nonG-guarantee}) may be interpreted as an indication that Gaussian noise is the worst one under a given variance constraint. It is interesting that a simple linear receiver suffices to provide such a performance guarantee.

In contrast, for quadratic receivers, there is no performance guarantee as (\ref{eqn:nonG-guarantee}). We can construct impulsive noise distributions such that the resulting ARPUC is arbitrarily close to zero. In fact, consider the mixed Gaussian distribution
\begin{eqnarray*}
P_\rvZ(\svz) = (1 - \epsilon) \frac{1}{\sqrt{\pi}}e^{-\svz^2} + \epsilon \frac{1}{\sqrt{\pi A}}e^{-\svz^2/A},
\end{eqnarray*}
where $A = (1 + \epsilon)/\epsilon > 1$ so $\mathrm{var}(Z) = 1$ for any $\epsilon \in (0, 1)$. The ARPUC (\ref{eqn:rate-nonG-q}) becomes (see Appendix \ref{app:nongaussian})
\begin{eqnarray}
\label{eqn:nongaussian-impulse}
\mathbf{R}_\mathrm{c}(\svx_\mathrm{c}, 1) = \sup_{\theta \geq 0} \frac{1}{\svx_\mathrm{c}^2}\left\{\theta (\svx_\mathrm{c}^2 + 1) + \log \sqrt{1 - \theta} - \log \left[1 + \epsilon (\sqrt{(1 - \theta)/(1 - \theta A)} - 1)\right]\right\},
\end{eqnarray}
for $0 < \theta < 1/A$. Therefore we can upper bound $\mathbf{R}_\mathrm{c}(\svx_\mathrm{c}, 1)$, upon noticing that the last two logarithmic terms in (\ref{eqn:nongaussian-impulse}) are both negative, by
\begin{eqnarray*}
\mathbf{R}_\mathrm{c}(\svx_\mathrm{c}, 1) &\leq& \sup_{0 < \theta < 1/A}\left\{\theta \left(1 + \frac{1}{\svx_\mathrm{c}^2}\right)\right\}\\
&=& \frac{1}{A} \left(1 + \frac{1}{\svx_\mathrm{c}^2}\right) \rightarrow 0,
\end{eqnarray*}
as $A \rightarrow \infty$, for arbitrary $\svx_\mathrm{c} \neq 0$. Intuitively, the quadratic receiver cannot attain robust performance against impulsive noise because it tends to ``amplify'' peaks in noise, which occur with a relatively high frequency for impulsive noise.
\end{example}

\subsection{Quadratic Receivers for Non-Coherent Fading Channels}

Consider the discrete-time memoryless fading channel
\begin{eqnarray}
\label{eqn:channel-fading}
\underline{\rvY} = \underline{\rvh}^\dag \underline{\rvx} + \underline{\rvZ},
\end{eqnarray}
where all the symbols are complex-valued. The input $\underline{\rvX}$ is $n_\mathrm{t}$-dimensional, the output $\underline{\rvY}$ and the additive noise $\underline{\rvZ}$ are $n_\mathrm{r}$-dimensional, and the fading matrix $\underline{\rvh}$ is $n_\mathrm{t} \times n_\mathrm{r}$. We assume that the noise is circularly symmetric complex Gaussian, {\it i.e.}, $\underline{\rvZ} \sim \mathcal{CN}(\underline{0}, \underline{I})$, and that the random fading matrix $\underline{\rvh}$ is independent of $\underline{\rvz}$ and has a distribution belonging to an uncertainty set, $\mathcal{P}_{\underline{\rvh}}$. When neither the transmitter nor the receiver has knowledge of the realizations of $\underline{\rvh}$, it is customary to employ quadratic receivers to process the channel output, {\it e.g.}, \cite{kailath60:it}, as
\begin{eqnarray}
g(\underline{\rvY}) = \underline{\rvY}^\dag \underline{G} \underline{\rvY},
\end{eqnarray}
where $\underline{G}$ is a deterministic positive semi-definite matrix. We define the cost function as the energy of an input, so that
\begin{eqnarray}
c(\underline{\rvX}) := \sum_{i = 1}^{n_\mathrm{t}} |\rvX_i|^2 = \|\underline{\rvX}\|^2,
\end{eqnarray}
and the zero-cost symbol is $\underline{\svx}_\mathrm{f} = \underline{0}$.

Following Proposition \ref{prop:rate}, the ARPUC can be shown to be
\begin{eqnarray}
\label{eqn:rate-fading}
\mathbf{R}_\mathrm{c}(\underline{\svx}_\mathrm{c}, \underline{G}) = \inf_{P_{\underline{\rvh}} \in \mathcal{P}_{\underline{\rvh}}} \sup_{\theta \geq 0} \frac{1}{\|\underline{\svx}_\mathrm{c}\|^2}\left\{
\mathrm{tr}[\underline{G}] \theta + \underline{\svx}_\mathrm{c}^\dag \mathbf{E}\left[\underline{\rvh} \underline{G} \underline{\rvh}^\dag \right] \underline{\svx}_\mathrm{c} \theta + \log \det(\underline{I} - \theta\underline{G})
\right\},
\end{eqnarray}
with the condition $\inf_{P_{\underline{\rvh}} \in \mathcal{P}_{\underline{\rvh}}} \underline{\svx}_\mathrm{c}^\dag \mathbf{E}\left[\underline{\rvh} \underline{G} \underline{\rvh}^\dag \right] \underline{\svx}_\mathrm{c} > 0$ satisfied. Note that $\theta$ is bounded such that $\det(\underline{I} - \theta\underline{G}) > 0$.

\begin{example}{\it (The Case of a Single Antenna at Either the Transmit or the Receive Side)}

Of particular simplicity is the case where the ARPUC only depends upon the covariance matrix of $\underline{\rvh}$. This property can hold when either the transmit side or the receive side has dimension one. In the multiple-input-single-output (MISO) case, there is no loss of generality by letting $G = 1$, and the ARPUC can be shown to be
\begin{eqnarray}
\mathbf{R}^{\mathrm{MISO}}_\mathrm{c}(\underline{\svx}_\mathrm{c}, 1) = \inf_{P_{\underline{\rvh}} \in \mathcal{P}_{\underline{\rvh}}} \frac{1}{\|\underline{\svx}_\mathrm{c}\|^2}\left\{
\underline{\svx}_\mathrm{c}^\dag \mathbf{E}(\underline{\rvh}\underline{\rvh}^\dag)\underline{\svx}_\mathrm{c}
- \log \left(1 + \underline{\svx}_\mathrm{c}^\dag \mathbf{E}(\underline{\rvh}\underline{\rvh}^\dag)\underline{\svx}_\mathrm{c} \right)
\right\}.
\end{eqnarray}

In the single-input-multiple-output (SIMO) case, the ARPUC becomes
\begin{eqnarray}
\label{eqn:simo}
\mathbf{R}^{\mathrm{SIMO}}_\mathrm{c}(\svx_\mathrm{c}, \underline{G}) = \inf_{P_{\underline{\rvh}} \in \mathcal{P}_{\underline{\rvh}}} \sup_{\theta \geq 0} \frac{1}{|\svx_\mathrm{c}|^2}\left\{ \mathrm{tr}[\mathbf{E}(\underline{\rvh}^\dag\underline{\rvh}) \underline{G}] |\svx_\mathrm{c}|^2 \theta + \mathrm{tr}[\underline{G}] \theta + \log\det(\underline{I} - \theta\underline{G})
\right\}.
\end{eqnarray}
Furthermore, if $\mathbf{E}(\underline{\rvh}^\dag\underline{\rvh}) = \underline{\Phi}$ for every $P_{\underline{\rvh}} \in \mathcal{P}_{\underline{\rvh}}$, we can simplify (\ref{eqn:simo}) to
\begin{eqnarray}
\label{eqn:simo-opt}
\mathbf{R}^{\mathrm{SIMO}}_\mathrm{c}(\svx_\mathrm{c}, \underline{G}) = \sup_{\theta \geq 0} \frac{1}{|\svx_\mathrm{c}|^2}\left\{
\mathrm{tr}[\underline{\Phi}\underline{G}]|\svx_\mathrm{c}|^2 \theta + \mathrm{tr}[\underline{G}] \theta + \log\det(\underline{I} - \theta\underline{G})
\right\}.
\end{eqnarray}
Maximizing (\ref{eqn:simo-opt}) with respect to $\theta$ and $\underline{G}$, we find that the optimal $\underline{G}$ is
\begin{eqnarray}
\underline{G}^\ast = \frac{1}{\theta} (\underline{I} + |\svx_\mathrm{c}|^2\underline{\Phi})^{-1} |\svx_\mathrm{c}|^2 \underline{\Phi},
\end{eqnarray}
and that the maximum ARPUC is
\begin{eqnarray}
\mathbf{R}^{\mathrm{SIMO}}_\mathrm{c}(\svx_\mathrm{c}, \underline{G}^\ast) = \mathrm{tr}[\underline{\Phi}] - \frac{\log\det(\underline{I} + |\svx_\mathrm{c}|^2 \underline{\Phi})}{|\svx_\mathrm{c}|^2},
\end{eqnarray}
which is identical to the CPUC when the fading matrix $\underline{\rvh}$ is circularly complex Gaussian and the channel input $\rvx$ has a peak constraint $|\svx_\mathrm{c}|$. The results of \cite{pierce66:it}, that orthogonal codes with energy detection achieve the CPUC for general fading Gaussian channels, correspond to a special case treated here, where both the input and the output are scalar, and the peak constraint $|\svx_\mathrm{c}| < \infty$ is removed.
\end{example}

\begin{example}{\it (Uniform Diagonal Quadratic Receiver)}

In this example, we restrict the weighing matrix $\underline{G}$ to be an identity matrix $\underline{I}_{n_\mathrm{r}\times n_\mathrm{r}}$. That is, the channel output processing is simply the sum of the squared magnitudes from individual receive antennas, but without any correlation among them. Such a structure is thus immune to possible phase offsets among receive antennas, and is easy to implement in antenna arrays.

The ARPUC (\ref{eqn:rate-fading}) now becomes
\begin{eqnarray}
\mathbf{R}_\mathrm{c}(\underline{\svx}_\mathrm{c}, \underline{I}_{n_\mathrm{r}\times n_\mathrm{r}}) = \inf_{P_{\underline{\rvh}} \in \mathcal{P}_{\underline{\rvh}}} \left\{
\frac{\underline{\svx}_\mathrm{c}^\dag \mathbf{E}[\underline{\rvh} \underline{\rvh}^\dag] \underline{\svx}_\mathrm{c}}{\|\underline{\svx}_\mathrm{c}\|^2} -
\frac{\log(1 + \underline{\svx}_\mathrm{c}^\dag \mathbf{E}[\underline{\rvh} \underline{\rvh}^\dag] \underline{\svx}_\mathrm{c}/n_\mathrm{r})}{\|\underline{\svx}_\mathrm{c}\|^2/n_\mathrm{r}}
\right\}.
\end{eqnarray}
It is apparent that the ARPUC only depends upon the fading covariance $\mathbf{E}[\underline{\rvh} \underline{\rvh}^\dag]$. If there is no peak constraint on $\|\underline{\svx}_\mathrm{c}\|^2$, the maximum ARPUC is achieved by letting $\|\underline{\svx}_\mathrm{c}\|^2 \rightarrow \infty$, and given by the following optimization problem:
\begin{eqnarray}
\label{eqn:optimization-fading}
\mathbf{R}_\mathrm{c} = \sup_{\tilde{\underline{\svx}}_\mathrm{c}: \|\tilde{\underline{\svx}}_\mathrm{c}\|^2 = 1} \inf_{P_{\underline{\rvh}} \in \mathcal{P}_{\underline{\rvh}}}
\tilde{\underline{\svx}}_\mathrm{c}^\dag \mathbf{E}[\underline{\rvh} \underline{\rvh}^\dag] \tilde{\underline{\svx}}_\mathrm{c}.
\end{eqnarray}
By utilizing the results of minimax robustness in Appendix \ref{app:general}, as done in Example \ref{example:gaussian-covariance}, we can find that, when the fading distribution uncertainty set $\mathcal{P}_{\underline{\rvh}}$ is compact and convex, the optimization (\ref{eqn:optimization-fading}) yields a saddle point, and its solution is $\mathbf{R}_\mathrm{c} = \min_{P_{\underline{\rvh}} \in \mathcal{P}_{\underline{\rvh}}} \lambda_{\max}(\mathbf{E}[\underline{\rvh} \underline{\rvh}^\dag])$.
\end{example}

\subsection{Photon-Counting Receivers for Poisson Channels}

We consider direct detection photon channels in which the channel observation can be modeled as point processes. For such channels without uncertainty, the capacity and error exponents have been fully identified; see \cite{davis80:it}, \cite{wyner88:it}, \cite{frey91:it}. For technical convenience, in this paper we adopt the discrete-time channel model as described in \cite[Example 2]{verdu90:it}, and concentrate on the case where the observation is a Poisson process with a fixed background photon flow rate, and that the channel input has no bandwidth constraint. For such a channel model, each ``channel use'' corresponds to a length-$T_0$ time duration, in which the transmitter modulates its input as a function $\rho(t)$, $t \in [0, T_0]$ such that the output is a Poisson point process with rate $\rho(t) + \rho_0$, where $\rho_0$ is the background photon flow rate. Therefore, an ideal photon-counting device within a channel use duration can detect a random number of photons with this number following a Poisson distribution of parameter $\left(\int_{0}^{T_0} \rho(t)dt + \rho_0 T_0\right)$. The cost associated with a channel input is equal to $\int_0^{T_0} \rho(t)dt$, therefore the zero-cost input symbol is $\rho(t) = 0$, $t \in [0, T_0]$. Let us model the channel state by allowing the fixed background photon flow rate to be uncertain, $\rho_0 \in [\underline{\rho_0}, \overline{\rho_0}]$. In a photon-counting receiver, the channel output processing function $g(\cdot)$ is nothing but the number of photons detected by the ideal photon-counting device within a channel use duration.

Following Proposition \ref{prop:rate}, we obtain that for any nonzero-cost input $\rho_\mathrm{c}(t)$, $t \in [0, T_0]$, the ARPUC for a photon-counting receiver is
\begin{eqnarray}
\label{eqn:rate-poisson}
\mathbf{R}_\mathrm{c}(\rho_\mathrm{c}(t)) &=& \inf_{\rho_0} \sup_{\theta \geq 0} \frac{1}{\Delta}\left\{
\theta (
\Delta + \rho_0
) - \rho_0 (e^\theta - 1)
\right\}\nonumber\\
&=& \inf_{\rho_0} \frac{1}{\Delta}\left\{
(
\Delta + \rho_0
) \left[\log(
\Delta + \rho_0
) - \log \rho_0\right] - \Delta
\right\}\nonumber\\
\label{eqn:rate-poisson-1}
&=& \frac{1}{\Delta}\left\{
(
\Delta + \overline{\rho_0}
) \left[\log(
\Delta + \overline{\rho_0}
) - \log \overline{\rho_0}\right] - \Delta
\right\},
\end{eqnarray}
where $\Delta := (1/T_0)\int_0^{T_0} \rho_\mathrm{c}(t)dt$. Since the right hand side of (\ref{eqn:rate-poisson-1}) is monotonically increasing with $\Delta$, if we have a peak rate constraint $\rho(t) \leq \rho_1$, the optimal nonzero-cost input function is $\rho_\mathrm{c}(t) = \rho_1$ for $t \in [0, T_0]$ and thus $\Delta = \rho_1$. Consequently the photon-counting receiver achieves
\begin{eqnarray}
\label{eqn:rate-poisson-2}
\mathbf{R}_\mathrm{c} = \left(1 + \overline{\rho_0}/{\rho_1}\right)\log\left(1 + \rho_1/\overline{\rho_0}\right) - 1.
\end{eqnarray}
Comparing \cite[Example 2]{verdu90:it}, the ARPUC (\ref{eqn:rate-poisson-2}) coincides with the CPUC when the channel state is $\rho = \overline{\rho_0}$, the most noisy one. So a photon-counting receiver indeed achieves the CPUC of the compound Poisson channel.

However, we note that, the photon-counting receiver may not attain robustness against non-Poisson point processes. Analogously to the impulsive additive-noise channels considered in Example \ref{example:non-Gaussian}, it is possible to construct impulsive point processes such that the resulting ARPUC with photon-counting receivers is arbitrarily close to zero. Consider a photon channel in which the background photon flow yields only two possible outputs at the ideal photon-counting device. Within a channel use, the device either detects $A$ photons with probability $\lambda_0/A$, or detects no photon, with probability $(1 - \lambda_0/A)$. The integer parameter $A >0$ may be made arbitrarily large. Note that the average rate of the background photon flow is $\lambda_0$.

Following Proposition \ref{prop:rate}, we obtain that the corresponding ARPUC is
\begin{eqnarray}
\label{eqn:non-poisson}
\mathbf{R}_\mathrm{c} = \sup_{\theta \geq 0}\frac{1}{\Delta}\left\{
\theta (\Delta + \lambda_0) - \log\left[
1 + \frac{\lambda_0}{A}(e^{\theta A} - 1)
\right]
\right\}.
\end{eqnarray}
Inspecting (\ref{eqn:non-poisson}), we notice that for any $\theta > 0$, as $A \rightarrow \infty$, the logarithmic term will eventually exceed the preceding term $\theta(\Delta + \lambda_0)$. Therefore as $A \rightarrow \infty$, we have to let $\theta \rightarrow 0$, and consequently the ARPUC vanishes asymptotically.

\section{CPUC Bounds for Channels with Convex Uncertainty Structure}
\label{sec:capacity}

In this section, we optimize the ARPUC in Proposition \ref{prop:rate} to quantify the ultimate performance of orthogonal codes. The optimization is analytically tractable when the uncertainty set of channel transition statistics satisfies a convexity property, by utilizing results of minimax robustness developed in \cite{verdu84:it}. To aid the reader, we briefly recapitulate the notations and basic results of minimax robustness in Appendix \ref{app:general}.

To avoid additional technicalities, throughout this section we focus on discrete and finite alphabets $\mathcal{X}$ and $\mathcal{Y}$, and assume that the cost function $c(\svx)$ is strictly positive and finite for every $\svx \in \mathcal{X}^\prime$. We note that sometimes it may be possible to extend the results to more general alphabets, while caution should be exercised to actually verify the corresponding technical conditions.

The following convexity property will be critical in this section.
\begin{defn}{\it (Convexity Property of Channel Uncertainty)}
\label{defn:convexity}
For any two channel states $s_1 \neq s_2$ in $\mathcal{S}$ and an arbitrary $\alpha \in [0, 1]$, there exists another state $s_\alpha \in \mathcal{S}$, such that $P_{s_\alpha}(\svy|\svx) = \alpha P_{s_1}(\svy|\svx) + (1 - \alpha) P_{s_2}(\svy|\svx)$ for every pair of $(\svx, \svy) \in \mathcal{X} \times \mathcal{Y}$.
\end{defn}

By this definition and the finiteness of the alphabet $\mathcal{Y}$, an immediate observation is that, for every $\svx \in \mathcal{X}$, the set of conditional channel output distributions, $\mathcal{P}_\svx := \{P_s(\cdot|\svx): s \in \mathcal{S}\}$, is compact and convex on $\mathcal{S}$.

\subsection{Some Information-Theoretic Results}
\label{subsec:info-theory}

Before proceeding with the ARPUC formula in Proposition \ref{prop:rate}, it is useful to conduct an information-theoretic analysis from the general CPUC formula of Proposition \ref{prop:general}. For the capacity game of compound channels, the convexity property we have assumed is sufficient to guarantee the existence of a saddle point, and consequently the capacity can be achieved by using a maximum-likelihood decoder designed for the saddle point channel realization \cite{stiglitz66:it, csiszar95:it, lapidoth98:it}. For the CPUC game of (\ref{eqn:cpuc-general}) in Proposition \ref{prop:general}, we can also establish a minimax result, as given by the following proposition.

\begin{prop}
\label{prop:cpuc}
For a compound memoryless channel $\mathbsf{M}_\mathrm{c} = \left(\mathcal{X}, \mathcal{Y}, \mathcal{S}, P_{\cdot}(\cdot|\cdot), c(\cdot)\right)$, in which $\mathcal{X}$ and $\mathcal{Y}$ are discrete and finite sets, consider the following two conditions:
\begin{itemize}
\item[(a)] there exists a unique zero-cost symbol $\svx_\mathrm{f}$ in the input alphabet $\mathcal{X}$;
\item[(b)] the convexity property in Definition \ref{defn:convexity}.
\end{itemize}
If (a) is satisfied, then the CPUC is
\begin{eqnarray}
\label{eqn:cpuc-lp}
\mathbf{C}_\mathrm{c} = \sup_{\stackrel{r(\svx): r(\svx) \geq 0}{\sum_{\svx \in \mathcal{X}^\prime} r(\svx) = 1}} \min_{s\in \mathcal{S}} \sum_{\svx \in \mathcal{X}^\prime} \frac{D\left(P_{s, \svx}(\svy)\|P_{s, \mathrm{f}}(\svy)\right)}{c(\svx)} r(\svx).
\end{eqnarray}
If (b) is satisfied, then the CPUC is
\begin{eqnarray}
\label{eqn:cpuc-minimax}
\mathbf{C}_\mathrm{c} = \min_{P_s(\cdot|\cdot): s\in \mathcal{S}} \sup_{\stackrel{\beta > 0, P_\rvX}{\mathbf{E}[c(\rvX)] \leq \beta}} \frac{I(\rvX; \rvY)}{\beta}.
\end{eqnarray}
Furthermore, if both (a) and (b) are satisfied, then we have
\begin{eqnarray}
\label{eqn:cpuc-zerocost}
\mathbf{C}_\mathrm{c} = \min_{s \in \mathcal{S}}
\sup_{\svx_\mathrm{c} \in \mathcal{X}^\prime} \frac{D\left(P_{s, \mathrm{c}}(\svy)\|P_{s, \mathrm{f}}(\svy)\right)}{c(\svx_\mathrm{c})}.
\end{eqnarray}
\end{prop}

Part of the proof of Proposition \ref{prop:cpuc} relies on a general minimax theorem in \cite{sion58:pjm}. We state the theorem in its following form taken from \cite[Sec. 5]{simons95:survey}:
\begin{thm}
\label{thm:sion}
Let $F$ be a convex subset of a linear topological space $\mathcal{F}$, $Q$ be a compact convex subset of a linear topological space $\mathcal{Q}$, and $U: F \times Q \rightarrow \mathbb{R}$ be upper semi-continuous on $F$ and lower semi-continuous on $Q$. Suppose that, (a) for all $q \in Q$ and $\lambda \in \mathbb{R}$, the level set $GE(\lambda, q) := \{f: f \in F, U(f, q) \geq \lambda\}$ is convex; and, (b) for all $f \in F$ and $\lambda \in \mathbb{R}$, the level set $LE(f, \lambda) := \{q: q \in Q, U(f, q) \leq \lambda\}$ is convex. Then
\begin{eqnarray}
\min_{q \in Q} \sup_{f \in F} U(f, q) = \sup_{f \in F} \min_{q \in Q} U(f, q).
\end{eqnarray}
\end{thm}
In Theorem \ref{thm:sion}, the property (a) is called {\it quasiconcavity}, and the property (b) is called {\it quasiconvexity}. These are generalizations of the conventional notions of concavity and convexity, respectively. For compound channel capacity problems, such generalizations are not necessary. However, they are required here for establishing the minimax result for the compound channel CPUC problem.

\noindent{\it Proof of Proposition \ref{prop:cpuc}:} 

{\it Proof of (\ref{eqn:cpuc-lp}):} We first prove that for any $r(\svx): r(\svx) \geq 0, \sum_{\svx \in \mathcal{X}^\prime} r(\svx) = 1$,
\begin{eqnarray*}
\min_{s \in \mathcal{S}} \sum_{\svx \in \mathcal{X}^\prime} \frac{D\left(P_{s, \svx}(\svy)\|P_{s, \mathrm{f}}(\svy)\right)}{c(\svx)} r(\svx)
\end{eqnarray*}
is an ARPUC. For this purpose, we start with the general formula (\ref{eqn:cpuc-general}) and expand $I(\rvx; \rvy)$ as \cite[Eqn. (10)]{verdu90:it}
\begin{eqnarray}
\label{eqn:proof-1}
I(\rvx; \rvy) = \sum_{\svx \in \mathcal{X}^\prime} P_\rvx(\svx) \cdot D\left(P_s(\svy|\svx)\|P_s(\svy|\svx_\mathrm{f})\right) - D\left(P_\rvy(\svy)\|P_s(\svy|\svx_\mathrm{f})\right).
\end{eqnarray}
As we let $P_\rvx(\svx) \rightarrow 0$ for all $\svx \in \mathcal{X}^\prime$, the average cost $\beta \rightarrow 0$ since $c(\svx) < \infty$, and the last term in the right hand side of (\ref{eqn:proof-1}) vanishes like $o(\beta)$, following the achievability proof of \cite[Theorem 3]{verdu90:it}. Hence we have
\begin{eqnarray}
\frac{I(\rvx;\rvy)}{\beta} \rightarrow \frac{1}{\sum_{\svx \in \mathcal{X}^\prime} P_\rvx(\svx)\cdot c(\svx)} \sum_{\svx \in \mathcal{X}^\prime} P_\rvx(\svx) \cdot D\left(P_s(\svy|\svx)\|P_s(\svy|\svx_\mathrm{f})\right)
\end{eqnarray}
asymptotically. Then the achievability readily follows as we choose $P_\rvx(\svx)$ to satisfy
\begin{eqnarray}
r(\svx) = \frac{P_\rvx(\svx)\cdot c(\svx)}{\sum_{\tilde{\svx} \in \mathcal{X}^\prime} P_\rvx(\tilde{\svx})\cdot c(\tilde{\svx})}
\end{eqnarray}
for every $\svx \in \mathcal{X}^\prime$.

We then prove that the CPUC expression in (\ref{eqn:cpuc-lp}) cannot be exceeded by any input distributions. For this purpose, we simply need to upper bound $I(\rvx; \rvy)$ by
\begin{eqnarray*}
I(\rvx; \rvy) \leq \sum_{\svx \in \mathcal{X}^\prime} P_\rvx(\svx) \cdot D\left(P_s(\svy|\svx)\|P_s(\svy|\svx_\mathrm{f})\right),
\end{eqnarray*}
according to (\ref{eqn:proof-1}), and the remaining part is analogous to the converse proof of \cite[Theorem 3]{verdu90:it}.

{\it Proof of (\ref{eqn:cpuc-minimax}):} We shall apply Theorem \ref{thm:sion}. Let us start with the general formula (\ref{eqn:cpuc-general}) of $\mathbf{C}_\mathrm{c}$,
\begin{eqnarray}
\mathbf{C}_\mathrm{c} &=& \sup_{\stackrel{\beta > 0, P_\rvX}{\mathbf{E}[c(\rvX)] \leq \beta}} \inf_{P_s(\cdot|\cdot): s\in \mathcal{S}} \frac{I(\rvX; \rvY)}{\beta}\nonumber\\
&=& \sup_{P_\rvX} \inf_{P_s(\cdot|\cdot): s\in \mathcal{S}} \frac{I(\rvX; \rvY)}{\mathbf{E}[c(\rvX)]}.
\end{eqnarray}
The set of input distributions is clearly convex, and by the convexity property in assumption, the set of channel transition distributions is compact and convex. Consider the continuity conditions. By the assumption that $c(\svx) < \infty$ is bounded away from zero for all $\svx \in \mathcal{X}^\prime$, $I(\rvX; \rvY)/\mathbf{E}[c(\rvX)]$ is continuous in $P_\rvX$ for any fixed $s \in \mathcal{S}$. By expanding the channel mutual information as $I(\rvX; \rvY) = \sum_{\svx\in\mathcal{X}} P(\svx) D\left(P_s(\svy|\svx)\|\sum_{\svx \in \mathcal{X}} P(\svx) P_s(\svy|\svx)\right)$, we have that $I(\rvX; \rvY)/\mathbf{E}[c(\rvX)]$ is lower semi-continuous in $P_s(\cdot|\cdot)$ \cite{csiszar81:book}, for any fixed $P_\rvX$. Finally consider the level sets. For fixed $P_\rvX$ and $\beta$, $I(\rvX; \rvY)/\beta$ is convex in $P_s(\cdot|\cdot)$ \cite{cover91:book}. For fixed $s \in \mathcal{S}$, we need to prove that the corresponding level set, $GE(\lambda, s) = \{\beta > 0, P_\rvX: I(\rvX; \rvY)/\beta \geq \lambda\}$, is convex. Fix $(\beta_1, P_\rvX^{(1)})$ and $(\beta_2, P_\rvX^{(2)})$ that belong to $GE(\lambda, s)$. Consider
\begin{eqnarray*}
\beta_\alpha &=& \alpha \beta_1 + (1 - \alpha) \beta_2,\\
P_\rvX^{(\alpha)} &=& \alpha P_\rvX^{(1)} + (1 - \alpha) P_\rvX^{(2)},
\end{eqnarray*}
for an arbitrary $\alpha \in [0, 1]$. Then we have
\begin{eqnarray*}
\frac{I_{P_\rvX^{(\alpha)}}(\rvX; \rvY)}{\beta_\alpha} &\geq& \frac{\alpha I_{P_\rvX^{(1)}}(\rvX; \rvY) + (1 - \alpha) I_{P_\rvX^{(2)}}(\rvX; \rvY)}{\beta_\alpha}\\
&\geq& \frac{\alpha \lambda \beta_1 + (1 - \alpha) \lambda \beta_2}{\beta} = \lambda,
\end{eqnarray*}
where the first inequality is from the concavity property of mutual information in input distributions \cite{cover91:book}, and the second inequality is from the definition of $GE(\lambda, s)$. In summary, we have utilized Theorem \ref{thm:sion} to establish the minimax equality (\ref{eqn:cpuc-minimax}).

{\it Proof of (\ref{eqn:cpuc-zerocost}):} Since we have established (\ref{eqn:cpuc-minimax}), the relationship (\ref{eqn:cpuc-zerocost}) directly follows from applying Proposition \ref{prop:matched}.
{\bf Q.E.D.}

In the following subsections, we return to the orthogonal coding scheme, to make comparison between its ultimate performance and the CPUC (\ref{eqn:cpuc-zerocost}).

\subsection{Maximum ARPUC of the Orthogonal Coding Scheme}
\label{subsec:lower-bound}

We start with the special case in which $\min_{s \in \mathcal{S}} D(P_{s, \mathrm{c}}(\svy)\|P_{s, \mathrm{f}}(\svy)) = \infty$. Since the output alphabet $\mathcal{Y}$ is finite, we can assert that, there exists an output $\bar{\svy} \in \mathcal{Y}$, such that $P_s(\bar{\svy}|\svx_\mathrm{c}) > 0$ and $p_s(\bar{\svy}|\svx_\mathrm{f}) = 0$ for all $s\in \mathcal{S}$. Therefore if we let the transformation function be $g(\svy) = 1$ if $\svy = \bar{\svy}$, and $g(\svy) = 0$ otherwise, the orthogonal coding scheme can achieve any arbitrarily large ARPUC. So in this case we have that the CPUC is $\mathbf{C}_\mathrm{c} = \infty$.

Now let us turn to the case where $\min_{s \in \mathcal{S}} D(P_{s, \mathrm{c}}(\svy)\|P_{s, \mathrm{f}}(\svy)) < \infty$. For the ARPUC formula in Proposition \ref{prop:rate}, by formally taking its supremum over $\svx_\mathrm{c} \in \mathcal{X}^\prime$ and $g(\cdot)$, we arrive at
\begin{eqnarray}
\label{eqn:rate-raw}
\mathbf{R}_\mathrm{c} = \sup_{\svx_\mathrm{c} \in \mathcal{X}^\prime, g(\cdot)} \inf_{s \in \mathcal{S}} \sup_{\theta \geq 0}\frac{1}{c(\svx_\mathrm{c})}\left\{
 \theta \mathbf{E}_{P_{s, \mathrm{c}}} [g(\rvY)] - \log \mathbf{E}_{P_{s, \mathrm{f}}} \left\{\exp[\theta g(\rvY)]\right\}
\right\}.
\end{eqnarray}

Since we have assumed the convexity of the uncertainty sets $\mathcal{P}_{\svx_\mathrm{c}}$ and $\mathcal{P}_{\svx_\mathrm{f}}$, we can utilize the minimax robustness results in \cite{verdu84:it} to verify that the order of $\sup_{g(\cdot)}$ and $\inf_{s \in \mathcal{S}}$ is interchangeable, as shown in Appendix \ref{app:interchange}. Thus,
\begin{eqnarray}
\label{eqn:rate-1}
\mathbf{R}_\mathrm{c} = \sup_{\svx_\mathrm{c} \in \mathcal{X}^\prime} \min_{s \in \mathcal{S}} \sup_{\theta \geq 0, g(\cdot)}
\frac{1}{c(\svx_\mathrm{c})}\left\{
\theta \mathbf{E}_{P_{s, \mathrm{c}}} [g(\rvY)] - \log \mathbf{E}_{P_{s, \mathrm{f}}} \left\{\exp[\theta g(\rvY)]\right\}
\right\}.
\end{eqnarray}

For fixed $\svx_\mathrm{c}$ and $(P_{s, \mathrm{f}}(\svy), P_{s, \mathrm{c}}(\svy))$, the inner supremum operator of (\ref{eqn:rate-1}) yields the following solution,
\begin{eqnarray}
\label{eqn:sup-g}
\sup_{\theta \geq 0, g(\cdot)}
\left\{
\theta \mathbf{E}_{P_{s, \mathrm{c}}} [g(\rvY)] - \log \mathbf{E}_{P_{s, \mathrm{f}}} \left\{\exp[\theta g(\rvY)]\right\}
\right\} = D\left(P_{s, \mathrm{c}}(\svy)\|P_{s, \mathrm{f}}(\svy)\right),
\end{eqnarray}
achieved by
\begin{eqnarray}
\label{eqn:optimum-g}
g(\svy) = \log\frac{P_{s, \mathrm{c}}(\svy)}{P_{s, \mathrm{f}}(\svy)},\quad\mathrm{and}\;\theta = 1,
\end{eqnarray}
as shown in Appendix \ref{app:optimal}.

So far we have not considered the condition (\ref{eqn:threshold-condition}) in Proposition \ref{prop:rate}, which now becomes
\begin{eqnarray}
\inf_{s\in \mathcal{S}} \left\{
\mathbf{E}_{P_{s, \mathrm{c}}}\left[\log\frac{P_{s, \mathrm{c}}(\rvY)}{P_{s, \mathrm{f}}(\rvY)}\right] - \mathbf{E}_{P_{s, \mathrm{f}}}\left[\log\frac{P_{s, \mathrm{c}}(\rvY)}{P_{s, \mathrm{f}}(\rvY)}\right]
\right\} &>& 0,\\
\label{eqn:condition-uncertainty}
i.e.,\; \inf_{s\in \mathcal{S}}\left\{
D\left(P_{s, \mathrm{c}}(\svy)\|P_{s, \mathrm{f}}(\svy)\right) + D\left(P_{s, \mathrm{f}}(\svy)\|P_{s, \mathrm{c}}(\svy)\right)
\right\} &>& 0,
\end{eqnarray}
and thus holds true whenever $\inf_{s\in \mathcal{S}} D\left(P_{s, \mathrm{c}}(\svy)\|P_{s, \mathrm{f}}(\svy)\right) > 0$.

In summary, we establish the maximum ARPUC as given by the following proposition.
\begin{prop}
\label{prop:uncertainty}
For a compound memoryless channel $\mathbsf{M}_\mathrm{c} = \left(\mathcal{X}, \mathcal{Y}, \mathcal{S}, P_{\cdot}(\cdot|\cdot), c(\cdot)\right)$, in which $\mathcal{X}$ and $\mathcal{Y}$ are discrete and finite sets, if:
\begin{itemize}
\item[(a)] there exists a unique zero-cost symbol $\svx_\mathrm{f}$ in the input alphabet;
\item[(b)] $\mathbsf{M}_\mathrm{c}$ satisfies the convexity property in Definition \ref{defn:convexity},
\end{itemize}
then the maximum ARPUC achievable by the orthogonal coding schemes described in Section \ref{sec:orthogonal} is the following lower bound to the CPUC:
\begin{eqnarray}
\label{eqn:cap-uncertainty}
\underline{\mathbf{C}}_\mathrm{c} := \sup_{\svx_\mathrm{c} \in \mathcal{X}^\prime} \min_{s \in \mathcal{S}}
\frac{D\left(P_{s, \mathrm{c}}(\svy)\|P_{s, \mathrm{f}}(\svy)\right)}{c(\svx_\mathrm{c})}.
\end{eqnarray}
\end{prop}

{\it Remark:} Let us denote the solution to the supinf game (\ref{eqn:cap-uncertainty}) by $\svx_\mathrm{c}^\ast$ and $(P_\mathrm{f}^\ast, P_\mathrm{c}^\ast)$. To achieve $\underline{\mathbf{C}}_\mathrm{c}$ in Proposition \ref{prop:uncertainty}, the orthogonal code uses nonzero-cost input symbol $\svx_\mathrm{c}^\ast$, and the corresponding output transformation is the log-likelihood ratio processing $g^\ast(\svy) = \log \left[P_\mathrm{c}^\ast(\svy)/P_\mathrm{f}^\ast(\svy)\right]$.\\

The CPUC lower bound $\underline{\mathbf{C}}_\mathrm{c}$ coincides with the CPUC (\ref{eqn:cpuc-zerocost}) when the game
\begin{eqnarray*}
\left(
\svx_\mathrm{c}\in\mathcal{X}^\prime, s\in\mathcal{S}, \frac{D(P_{s, \mathrm{c}}(\svy)\|P_{s, \mathrm{f}}(\svy))}{c(\svx_\mathrm{c})}
\right)
\end{eqnarray*}
has an equilibrium. For such a purpose, it is possible to utilize either the various forms of minimax theorems (see, {\it e.g.}, \cite{simons95:survey} and references therein), or minimax robustness results \cite{verdu84:it}, depending upon the specific channel model encountered.

A comment regarding the computation of $\underline{\mathbf{C}}_\mathrm{c}$ follows. This KL-distance type CPUC lower bound is in principle simpler to compute than the general CPUC given in Proposition \ref{prop:general}. However, its computation may still be numerically non-trivial due to the hindrance posed by the requirement of convexity. For practical purposes, a compound channel model is often described by a certain class of distributions with a few unknown parameters, for example, additive Gaussian noise with an unknown variance. Even if such unknown parameters are from convex sets, the resulting channel transition distributions typically are not so. Therefore, the requirement of convexity in Definition \ref{defn:convexity} usually can only be fulfilled by additional convex-hull operations (see, {\it e.g.}, \cite{barvinok02:book}) upon the original channel uncertainty sets, inevitably leading to mixed probability distributions, for which closed-form expressions for the KL distance rarely exist.

Here we give an example to illustrate that the CPUC lower bound can be strictly smaller than the CPUC.

\begin{example}
\label{example:bsc}
Consider a compound channel with three possible inputs $\{a, b, f\}$ with costs $c(a) = c(b) = 1$ and $c(f) = 0$ (thus $\svx_\mathrm{f} = f$). For each input symbol the possible outputs are $\{a', b'\}$. The conditional output distributions $P(\cdot|a)$ and $P(\cdot|b)$ are unique without any uncertainty, given by
\begin{eqnarray*}
P(a'|a) = P(b'|b) = 1 - q, \quad P(b'|a) = P(a'|b) = q,
\end{eqnarray*}
where $0 \leq q \leq 1/2$ is a deterministic number. The conditional output distribution $P(\cdot|f)$ is contained by the convex hull of $P(\cdot|a)$ and $P(\cdot|b)$, represented by $P(\cdot|f) \in \{(P(a'|f), P(b'|f)) = (\delta, 1 - \delta): q \leq \delta \leq 1 - q\}$. In view of $\underline{\mathbf{C}}_\mathrm{c}$, we see that for either $\svx_\mathrm{c} = a$ or $b$, it is possible for $P_\mathrm{f}$ to coincide with $P_\mathrm{c}$. Therefore we have $\underline{\mathbf{C}}_\mathrm{c} = 0$. On the other hand, for every possible realization of $P(\cdot|f)$, we can choose $\svx_\mathrm{c}$ from $\{a, b\}$ such that $P_\mathrm{c} \neq P(\cdot|f)$. Consequently, we can calculate from (\ref{eqn:cpuc-zerocost}) that $\mathbf{C}_\mathrm{c} = \log 2 + q\log q + (1 - q)\log(1 - q) > 0$ for $0 \leq q < 1/2$. Indeed it is easily seen that $\mathbf{C}_\mathrm{c}$ can be achieved by coding over $\{a, b\}$ only, ignoring the existence of the free symbol $f$.
\end{example}

\subsection{Mixed Strategies Achieve the CPUC}
\label{subsec:mixed}

\begin{figure}[ht]
\centerline{\includegraphics[scale=0.7]{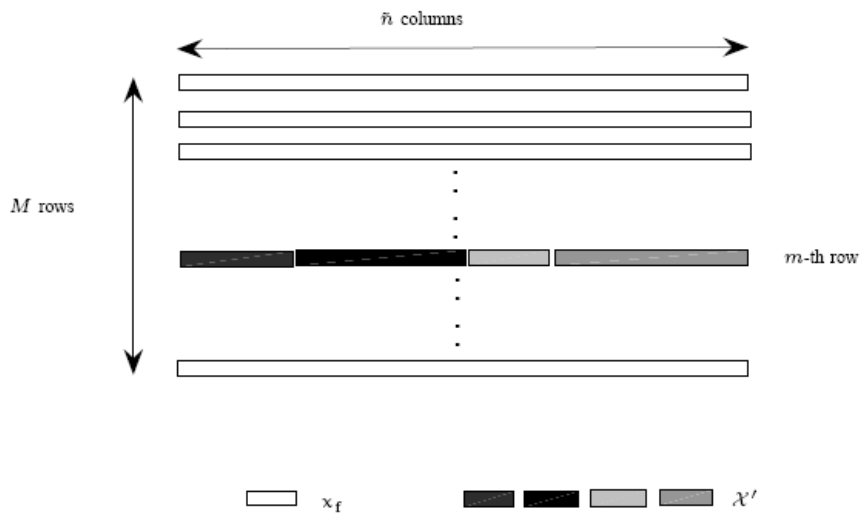}}
\caption{Illustration of an orthogonal codeword with mixed strategy.}
\label{fig:orthocodemix}
\end{figure}

The reader may observe from (\ref{eqn:cpuc-lp}) in Proposition \ref{prop:cpuc} as well as Example \ref{example:bsc} that it may be insufficient to use orthogonal codes with only one value of nonzero-cost symbols, because possibly for each nonzero-cost symbol value, the channel uncertainty set has a small (or even vanishing) minimum KL distance to the uncertainty set of the zero-cost symbol. Potential performance improvement, therefore, can be available by extending the orthogonal coding scheme to include different values of nonzero-cost symbols, in hope that for any possible channel realization there exist some ``good'' nonzero-cost inputs. To this end, we introduce the mixed strategy as follows (see Figure \ref{fig:orthocodemix}).

A mixed strategy is parametrized by a function $r(\svx) \geq 0$, $\svx \in \mathcal{X}^\prime$, satisfying $\sum_{\svx \in \mathcal{X}^\prime} r(\svx) = 1$. Thus $r(\cdot)$ may be interpreted as a probability mass function on $\mathcal{X}^\prime$. For an orthogonal coding scheme that uses the mixed strategy, to represent message $m$, the $\tilde{n}$ elements in the $m$th row of the codeword take $|\mathcal{X}^\prime|$ possible values. Specifically, $\tilde{n} \cdot r(\svx)$ of the $\tilde{n}$ elements take symbol value $\svx$, for every $\svx \in \mathcal{X}^\prime$.\footnote{Without loss of generality, by letting $\tilde{n}$ be sufficiently large, we assume that $\tilde{n}\cdot r(\svx)$ is integer-valued for all $\svx\in \mathcal{X}^\prime$.} All the $(M - 1)$ rows other than the $m$th row are all $\svx_\mathrm{f}$, as in the orthogonal coding scheme without mixed strategy described in Section \ref{sec:orthogonal}. For notational convenience, we denote the set of column indexes which correspond to input symbol $\svx$ by $\mathcal{I}_\svx$. It is obvious that $\bigcup_{\svx \in \mathcal{X}^\prime} \mathcal{I}_\svx = \{1, \ldots, \tilde{n}\}$, and $\mathcal{I}_\svx \cap \mathcal{I}_{\svx^\prime} = \emptyset$ for any $\svx^\prime \neq \svx$. The decoder knows the sets $\mathcal{I}_\cdot$.

The decoding algorithm is a slight modification of that in Section \ref{sec:orthogonal}. For each $\svx \in \mathcal{X}^\prime$, the decoder chooses a processing transformation function $g_\svx(\svy)$ for the channel outputs whose column indexes in the orthogonal codeword block belong to $\mathcal{I}_\svx$. The receiver then computes the decoding metrics as
\begin{eqnarray}
\rvt_m = \frac{1}{\tilde{n}}\cdot \sum_{\svx \in \mathcal{X}^\prime} \sum_{i \in \mathcal{I}_\svx} g_\svx(\rvy_{m, i}), \quad m = 1, \ldots, M.
\end{eqnarray}
The decoding rule is to declare the decoded message $\hat{m}$ as $\hat{m} = \mathrm{arg} \max_{m = 1, \ldots, M} \rvt_m$, the same as that in Section \ref{sec:orthogonal}.

Equipped with the convexity property in Definition \ref{defn:convexity}, orthogonal codes with the mixed strategy actually can be sufficient to achieve the CPUC of compound memoryless channels, as established by the following proposition.
\begin{prop}
\label{prop:rate-mixed}
For a compound memoryless channel $\mathbsf{M}_\mathrm{c} = \left(\mathcal{X}, \mathcal{Y}, \mathcal{S}, P_\cdot(\cdot|\cdot), c(\cdot)\right)$, in which $\mathcal{X}$ and $\mathcal{Y}$ are discrete and finite sets, if:
\begin{itemize}
\item[(a)] there exists a unique zero-cost symbol $\svx_\mathrm{f}$ in the input alphabet;
\item[(b)] $\mathbsf{M}_\mathrm{c}$ satisfies the convexity property in Definition \ref{defn:convexity},
\end{itemize}
then the maximum ARPUC achievable by the orthogonal coding schemes with mixed strategy coincides with the CPUC as given by (\ref{eqn:cpuc-zerocost}).
\end{prop}

{\it Proof:} Following the same line of steps in establishing Propositions \ref{prop:rate} and \ref{prop:uncertainty}, we can express the maximum ARPUC of orthogonal codes with mixed strategy as
\begin{eqnarray}
\label{eqn:cap-lb-mixed}
\underline{\mathbf{C}}_\mathrm{c, m} := \sup_{\stackrel{r(\svx): r(\svx) \geq 0}{\sum_{\svx \in \mathcal{X}^\prime} r(\svx) = 1}} \min_{s \in \mathcal{S}}
\frac{ \sum_{\svx \in \mathcal{X}^\prime} {D}\left(P_{s, \svx}(\svy)\|P_{s, \mathrm{f}}(\svy)\right) r(\svx)}{\sum_{\svx \in \mathcal{X}^\prime} c(\svx)r(\svx)}.
\end{eqnarray}
The subsequent proof again relies on Theorem \ref{thm:sion}. First, the mixed strategies $r(\cdot)$ clearly constitute a convex set, and by the convexity property in assumption, the set of channel transition distributions $\mathcal{P}_\svx$ is compact and convex. Second, the payoff function in (\ref{eqn:cap-lb-mixed}) is continuous in $r(\cdot)$, and lower semi-continuous on the set of channel transition distributions \cite{csiszar81:book}. Finally, for a fixed mixed strategy $r(\cdot)$, the payoff function is convex in the channel transition distributions \cite{cover91:book}. On the other hand, for a fixed channel realization $s \in \mathcal{S}$, the corresponding level set,
\begin{eqnarray*}
GE(\lambda, s) &=& \left\{r(\cdot): \frac{ \sum_{\svx \in \mathcal{X}^\prime} {D}\left(P_{s, \svx}(\svy)\|P_{s, \mathrm{f}}(\svy)\right) r(\svx)}{\sum_{\svx\in\mathcal{X}^\prime} c(\svx)r(\svx)} \geq \lambda
\right\}\\
&=& \left\{r(\cdot): \sum_{\svx\in \mathcal{X}^\prime} \left[ {D}\left(P_{s, \svx}(\svy)\|P_{s, \mathrm{f}}(\svy)\right) - \lambda c(\svx)
\right] r(\svx) \geq 0
\right\}
\end{eqnarray*}
is obviously a convex set. Therefore by utilizing Theorem \ref{thm:sion}, we have
\begin{eqnarray*}
\underline{\mathbf{C}}_\mathrm{c, m} &=& \min_{s \in \mathcal{S}} \sup_{\stackrel{r(\svx): r(\svx) \geq 0}{\sum_{\svx\in\mathcal{X}^\prime} r(\svx) = 1}}
\frac{ \sum_{\svx\in\mathcal{X}^\prime} {D}\left(P_{s, \svx}(\svy)\|P_{s, \mathrm{f}}(\svy)\right) r(\svx)}{\sum_{\svx\in\mathcal{X}^\prime} c(\svx)r(\svx)}\\
&=& \min_{s \in \mathcal{S}} \sup_{\svx_\mathrm{c} \in \mathcal{X}^\prime} \frac{{D}\left(P_{s, \mathrm{c}}(\svy)\|P_{s, \mathrm{f}}(\svy)\right)}{c(\svx_\mathrm{c})} = \mathbf{C}_\mathrm{c}
\end{eqnarray*}
as given by Proposition \ref{prop:cpuc}.
{\bf Q.E.D.}

{\it Example \ref{example:bsc} (cont.):} Now let us consider using mixed strategies for the channel in Example \ref{example:bsc}. It can be verified that the solution to (\ref{eqn:cap-lb-mixed}) is $r^\ast(a) = r^\ast(b) = 1/2$, $(P^\ast(a'|f), P^\ast(b'|f)) = (1/2, 1/2)$, and $\underline{\mathbf{C}}_\mathrm{c, m} = \log 2 + q\log q + (1 - q)\log (1 - q)$, which is the same as the CPUC $\mathbf{C}_\mathrm{c}$ obtained earlier. Therefore, by mixing up two nonzero-cost input symbols, we can achieve the CPUC of this compound channel.

\section{Conclusions}
\label{sec:conclusion}

For ergodic memoryless channels, when a zero-cost symbol exists in the input alphabet, an orthogonal code is sufficient to asymptotically achieve the CPUC. For compound memoryless channels, however, the ignorance of channel state realization generally prevents the orthogonal coding scheme from being optimal. By extending the orthogonal decoding algorithm for ergodic memoryless channels, we obtain a class of ARPUC for compound memoryless channels. The utility of this class of ARPUC is illustrated by analyzing several practical receivers for several representative channels. In this paper, we specifically study linear and quadratic receivers for linear additive-noise channels, quadratic receivers for non-coherent fading channels, and photon-counting receivers for Poisson channels.

Under the condition that the uncertainty set of channel transition statistics satisfies a certain convex property, we systematically optimize the performance of orthogonal codes to obtain a lower bound to the CPUC which involves the KL distance between two conditional output distributions, as well as a minimax game between selecting the optimal nonzero-cost input symbol and selecting the least favorable channel state realization. The CPUC lower bound achieved by an orthogonal code without mixed strategy is tight if the minimax game has an equilibrium. Moreover, we propose a way to extend the orthogonal coding scheme, by allowing a mixed strategy in which an orthogonal code contains an appropriate composition of different nonzero-cost input symbols. Such a mixed strategy improves the ARPUC of orthogonal codes without mixed strategy, and its optimization actually leads to the CPUC, under the convexity condition.

In closing, we briefly comment on some open issues unaddressed in this paper. First, regarding the results obtained, it would be desirable to lift some of the technical conditions for further generality. Specifically, the assumption of convexity that we utilize in Section \ref{sec:capacity} seems to be a crucial prerequisite, but as commented, it also considerably limits the applicability of the resulting ARPUC and CPUC. For example, the results in Section \ref{sec:capacity} cannot replace the {\it ad hoc} analysis in Section \ref{sec:application}, because those channel models examined there generally do not satisfy the convexity condition. Second, as we view the development in this paper as an initial step toward a full understanding of the robustness issue for wideband communication systems, it would be useful to examine more bandwidth-efficient coding schemes other than orthogonal codes. Despite its simplicity, the orthogonal coding scheme suffers from the slow (sub-exponential) growth rate of the number of messages with the coding block length, implying its extremely low bandwidth efficiency. To this end, it will be of considerable interest to examine and compare the wideband slopes \cite{verdu02:it} of different coding schemes for compound channels.

\section{Appendix}
\label{sec:appendix}

\subsection{Proof of Proposition \ref{prop:general}}
\label{app:proof-general}

We first show the achievability of $\mathbf{C}_\mathrm{c}$. The compound channel coding theorem \cite[Theorem 1]{blackwell59:ams} states that, for every $\beta > 0$ and every $0 < \epsilon < 1$, the rate $C_\mathrm{c}(\beta) := \sup_{P_\rvx: \mathbf{E}[c(\rvx)] \leq \beta} \inf_{P_s(\cdot|\cdot):s\in \mathcal{S}} I(\rvx; \rvy)$ is $\epsilon$-achievable under average cost $\beta$. Therefore, for every fixed $\gamma > 0$, there exists $n_\beta$ such that if $n > n_\beta$, then an $(n, M, n\beta, \epsilon)$ code can be found with
\begin{eqnarray}
\frac{\log M}{n} > C_\mathrm{c}(\beta) - \frac{\gamma\beta}{2}.
\end{eqnarray}
The subsequent arguments of the achievability then directly follow \cite[Theorem 2]{verdu90:it}.

We then show the converse of $\mathbf{C}_\mathrm{c}$. From \cite[Lemma 6]{blackwell59:ams}, we obtain that for every possible channel realization $s \in \mathcal{S}$, every $(n, M, \nu, \epsilon)$ code should satisfy
\begin{eqnarray}
\frac{\log M}{\nu} \leq \frac{1}{1 - \epsilon}\left\{
\frac{n}{\nu} I(\rvx; \rvy) + \frac{\log 2}{\nu}
\right\}.
\end{eqnarray}
Considering the least favorable channel realization, we have
\begin{eqnarray}
\frac{\log M}{\nu} \leq \frac{1}{1 - \epsilon}\left\{\frac{n}{\nu} \inf_{P_s(\cdot|\cdot):s \in \mathcal{S}}I(\rvx; \rvy) + \frac{\log 2}{\nu}\right\}.
\end{eqnarray}
Furthermore, optimizing the distribution of $\rvx$ under the average cost constraint like in \cite[proof of Theorem 2]{verdu90:it}, we obtain
\begin{eqnarray}
\frac{\log M}{\nu} \leq \frac{1}{1 - \epsilon}\left\{\sup_{\beta > 0} \frac{1}{\beta} \sup_{\rvx: \mathbf{E}[c(\rvx)] \leq \beta}\inf_{P_s(\cdot|\cdot):s \in \mathcal{S}}I(\rvx; \rvy) + \frac{\log 2}{\nu}\right\}.
\end{eqnarray}
So if $\mathbf{R}_\mathrm{c}$ is $\epsilon$-achievable per unit cost, then for every $\gamma > 0$, there exists $\nu_0$ such that for any $\nu > \nu_0$
\begin{eqnarray}
\label{eqn:converse-1}
\mathbf{R}_\mathrm{c} - \gamma < \frac{1}{1 - \epsilon} \left\{
\sup_{\stackrel{\beta > 0, P_\rvx}{\mathbf{E}[c(\rvx)] \leq \beta}} \inf_{P_s(\cdot|\cdot):s\in\mathcal{S}} \frac{I(\rvx; \rvy)}{\beta} + \frac{\log 2}{\nu}
\right\}.
\end{eqnarray}
By letting $\gamma \rightarrow 0$, $\epsilon \rightarrow 0$, and $\nu \rightarrow \infty$, we have
\begin{eqnarray}
\mathbf{R}_\mathrm{c} \leq \sup_{\stackrel{\beta > 0, P_\rvx}{\mathbf{E}[c(\rvx)] \leq \beta}} \inf_{P_s(\cdot|\cdot):s\in\mathcal{S}} \frac{I(\rvx; \rvy)}{\beta},
\end{eqnarray}
and thus the converse is established.

\subsection{On the Tightness of the ARPUC in Proposition \ref{prop:rate}}
\label{app:arpuc-max}

In this appendix we show that the ARPUC in Proposition \ref{prop:rate} is also the maximum ARPUC for the given orthogonal coding scheme. To this end, it suffices to show that for every channel state $s \in \mathcal{S}$, the decoding algorithm cannot achieve arbitrarily small error probability if the growth of $M$ satisfies
\begin{eqnarray}
\label{eqn:growth-M-bad}
\frac{\log M}{\tilde{n}} = \sup_{\theta \geq 0} \left[
\theta \mathbf{E}_{P_{s, \mathrm{c}}}[g(\rvY)] - \log \mathbf{E}_{P_{s, \mathrm{f}}}[\exp(\theta g(\rvY))]
\right] + \delta,
\end{eqnarray}
as $\tilde{n}\rightarrow \infty$, for any arbitrarily small $\delta > 0$.

For $\max_{m = 2, \ldots, M} \rvt_m$, consider the following lower bound as
\begin{eqnarray}
\label{eqn:max-lb}
&&\mathrm{Pr}\left[\max_{m = 2, \ldots, M} \rvt_m > \mathbf{E}_{P_{s, \mathrm{c}}}[g(\rvY)]\right]\nonumber\\
&=& 1 - \mathrm{Pr}\left[\max_{m = 2, \ldots, M} \rvt_m \leq \mathbf{E}_{P_{s, \mathrm{c}}}[g(\rvY)]\right]\nonumber\\
&=& 1 - \left\{1 - \mathrm{Pr}\left[\rvt_2 > \mathbf{E}_{P_{s, \mathrm{c}}}[g(\rvY)]\right]\right\}^{M - 1}\nonumber\\
&\geq& 1 - \exp\left\{-(M - 1)\mathrm{Pr}\left[\rvt_2 > \mathbf{E}_{P_{s, \mathrm{c}}}[g(\rvY)]\right]\right\},
\end{eqnarray}
where the inequality is from $(1 - t)^{M - 1} \leq \exp[-(M - 1)t]$ for $t \in [0, 1]$. By Cram\'{e}r's theorem (see, {\it e.g.}, \cite[Theorem 2.2.3]{dembo98:book}), the tail probability $\mathrm{Pr}\left[\rvt_2 > \mathbf{E}_{P_{s, \mathrm{c}}}[g(\rvY)]\right]$ scales as
\begin{eqnarray}
\label{eqn:cramer}
{\liminf}_{\tilde{n}\rightarrow \infty} \frac{1}{\tilde{n}} \log \mathrm{Pr}\left[\rvt_2 > \mathbf{E}_{P_{s, \mathrm{c}}}[g(\rvY)]\right] \geq - \inf_{t: t > \mathbf{E}_{P_{s, \mathrm{c}}}[g(\rvY)]} \sup_{\theta \in \mathbb{R}} \left\{\theta t - \log \mathbf{E}_{P_{s, \mathrm{f}}}[\exp(\theta g(\rvY))]\right\}.
\end{eqnarray}
Under the additional technical condition that $\log \mathbf{E}_{P_{s, \mathrm{f}}}[\exp(\theta g(\rvY))] < \infty$ for some $\theta > 0$, the bound (\ref{eqn:cramer}) further reduces into (see, {\it e.g.}, \cite[Lemma 2.2.5]{dembo98:book})
\begin{eqnarray}
\label{eqn:cramer-1}
{\liminf}_{\tilde{n}\rightarrow \infty} \frac{1}{\tilde{n}} \log \mathrm{Pr}\left[\rvt_2 > \mathbf{E}_{P_{s, \mathrm{c}}}[g(\rvY)]\right] \geq - \sup_{\theta \geq 0} \left\{\theta \mathbf{E}_{P_{s, \mathrm{c}}}[g(\rvY)] - \log \mathbf{E}_{P_{s, \mathrm{f}}}[\exp(\theta g(\rvY))]\right\}.
\end{eqnarray}
Therefore, if the growth of $M$ satisfies (\ref{eqn:growth-M-bad}) for any $\delta > 0$, from (\ref{eqn:max-lb}) and (\ref{eqn:cramer-1}) we have
\begin{eqnarray}
\mathrm{Pr}\left[\max_{m = 2, \ldots, M} \rvt_m > \mathbf{E}_{P_{s, \mathrm{c}}}[g(\rvY)]\right] \rightarrow 1
\end{eqnarray}
as $\tilde{n} \rightarrow \infty$. This establishes the tightness of the ARPUC in Proposition \ref{prop:rate}.

\subsection{Proof of (\ref{eqn:gaussian-covariance})}
\label{app:gaussian-covariance}

Our proof of (\ref{eqn:gaussian-covariance}) hinges on the theory of minimax robustness developed in \cite{verdu84:it}. Since this theory is also of a central technical role in Section \ref{sec:capacity}, to facilitate the reader, we briefly recapitulate notation and key results from \cite{verdu84:it} in Appendix \ref{app:general}. In the context of Example \ref{example:gaussian-covariance}, $\underline{w}_\mathrm{l}$ is the filter, $\underline{\Phi} \in \mathcal{S}_{\underline{\Phi}}$ is the operating point, and $1/(\underline{w}_\mathrm{l}^\mathrm{T} \underline{\Phi}\underline{w}_\mathrm{l})$ is the payoff function. We will show that the game has a saddle point, such that
\begin{eqnarray}
\mathbf{R}_\mathrm{c} &=& \sup_{\|\underline{w}_\mathrm{l}\| = 1} \inf_{\underline{\Phi} \in \mathcal{S}_{\underline{\Phi}}} \frac{1}{2\underline{w}_\mathrm{l}^\mathrm{T}\underline{\Phi}\underline{w}_\mathrm{l}}\nonumber\\
&=& \min_{\underline{\Phi} \in \mathcal{S}_{\underline{\Phi}}} \sup_{\|\underline{w}_\mathrm{l}\| = 1} \frac{1}{2\underline{w}_\mathrm{l}^\mathrm{T}\underline{\Phi}\underline{w}_\mathrm{l}}\nonumber\\
\label{eqn:app-temp1}
&=& \frac{1}{2 \max_{\underline{\Phi} \in \mathcal{S}_{\underline{\Phi}}} \lambda_{\min}(\underline{\Phi})}.
\end{eqnarray}

First, by assumption, $\mathcal{S}_{\underline{\Phi}}$ is convex, and for every $\underline{w}_\mathrm{l}$, the payoff function $1/(\underline{w}_\mathrm{l}^\mathrm{T} \underline{\Phi}\underline{w}_\mathrm{l})$ is convex with respect to $\underline{\Phi}$, due to the convexity of function $f(t) = 1/t$.

The least favorable operating point is easily shown from (\ref{eqn:app-temp1}) to be the covariance matrix $\underline{\Phi}^\ast \in \mathcal{S}_{\underline{\Phi}}$ that maximizes $\lambda_{\min}(\underline{\Phi})$, and the corresponding optimal filter $\underline{w}_\mathrm{l}^\ast$ is the associated unit-norm eigenvector.

To complete the proof, we need to show that $(\underline{w}_\mathrm{l}^\ast, \underline{\Phi}^\ast)$ is a regular pair. To this end, for every $\underline{\Phi} \in \mathcal{S}_{\underline{\Phi}}$, consider the neighboring operating point $\underline{\Phi}_\alpha = (1 - \alpha) \underline{\Phi}^\ast + \alpha \underline{\Phi}$ for small $\alpha \in [0, 1]$. On one hand, the payoff function maximized over $\{\underline{w}_\mathrm{l}: \|\underline{w}_\mathrm{l}\| = 1\}$ for a fixed operating point $\underline{\Phi}_\alpha$ is $1/\lambda_{\min}(\underline{\Phi}_\alpha)$. By utilizing the matrix perturbation theory ({\it e.g.}, \cite{kato76:book}), this payoff value behaves like
\begin{eqnarray}
\label{eqn:app-temp2}
\frac{1}{\lambda_{\min}(\underline{\Phi}_\alpha)} = \frac{1}{(1 - \alpha)\lambda_{\min}(\underline{\Phi}^\ast) + \alpha (\underline{w}_\mathrm{l}^\ast)^\mathrm{T} \underline{\Phi} \underline{w}_\mathrm{l}^\ast + o(\alpha)}
\end{eqnarray}
as $\alpha \rightarrow 0$. On the other hand, the payoff function for $(\underline{w}_\mathrm{l}^\ast, \underline{\Phi}_\alpha)$ is
\begin{eqnarray}
\frac{1}{(\underline{w}_\mathrm{l}^\ast)^\mathrm{T} \underline{\Phi}_\alpha \underline{w}_\mathrm{l}^\ast} &=& \frac{1}{(1 - \alpha) (\underline{w}_\mathrm{l}^\ast)^\mathrm{T} \underline{\Phi}^\ast \underline{w}_\mathrm{l}^\ast + \alpha (\underline{w}_\mathrm{l}^\ast)^\mathrm{T} \underline{\Phi} \underline{w}_\mathrm{l}^\ast}\nonumber\\
\label{eqn:app-temp3}
&=& \frac{1}{(1 - \alpha)\lambda_{\min}(\underline{\Phi}^\ast) + \alpha (\underline{w}_\mathrm{l}^\ast)^\mathrm{T} \underline{\Phi} \underline{w}_\mathrm{l}^\ast}.
\end{eqnarray}
Comparing (\ref{eqn:app-temp2}) and (\ref{eqn:app-temp3}), we find that their difference scales like $o(\alpha)$ as $\alpha \rightarrow 0$, if $\lambda_{\min}(\underline{\Phi}^\ast) = \sup_{\underline{\Phi}\in \mathcal{S}_{\underline{\Phi}}}\lambda_{\min}(\underline{\Phi}) > c$ for some finite positive $c > 0$. Hence $(\underline{w}_\mathrm{l}^\ast, \underline{\Phi}^\ast)$ is a regular pair, and by the saddle point property we establish the validity of (\ref{eqn:gaussian-covariance}).

\subsection{Proof of (\ref{eqn:nongaussian-impulse})}
\label{app:nongaussian}

From (\ref{eqn:rate-nonG-q}) we get
\begin{eqnarray}
\mathbf{R}_\mathrm{c}(\svx_\mathrm{c}, 1) = \sup_{\theta \geq 0} \frac{1}{\svx_\mathrm{c}^2} \left\{\theta(\svx_\mathrm{c}^2 + 1) - \log \mathbf{E}[\exp(\theta \rvz^2)]\right\}.
\end{eqnarray}
For the mixed Gaussian noise, We can evaluate $\mathbf{E}[\exp(\theta \rvz^2)]$ as
\begin{eqnarray}
\mathbf{E}[\exp(\theta \rvz^2)] &=& \frac{(1 - \epsilon)}{\sqrt{\pi}}\int_\mathbb{R} e^{-(1 - \theta)\svz^2} d\svz + \frac{\epsilon}{\sqrt{\pi A}} \int_\mathbb{R} e^{-(1/A - \theta)\svz^2}d\svz\nonumber\\
&=& \frac{1 - \epsilon}{\sqrt{1 - \theta}} + \frac{\epsilon}{\sqrt{1 - A\theta}},
\end{eqnarray}
if and only if $\theta < 1/A < 1$. Then (\ref{eqn:nongaussian-impulse}) follows from direct manipulations.

\subsection{Summary of Notation and Key Results of Minimax Robustness}
\label{app:general}

The material of this subsection is from \cite{verdu84:it}, and some changes in notation are adopted to meet the notational convention of the current paper.

Denote by $\mathcal{F}$ and $\mathcal{Q}$ two linear topological spaces, called the space of filters and the space of operating points, respectively. The payoff, or utility, function $U$ is a real functional
\begin{eqnarray}
U: \mathcal{F} \times \mathcal{Q} \rightarrow \mathbb{R}.
\end{eqnarray}
Suppose that $F \subset \mathcal{F}$ is the set of allowable filters and $Q \subset \mathcal{Q}$ is the set of possible operating points. Let us refer to the triple $(F, Q, U)$ as a game, in which $U$ is maximized over $F$ and minimized over $Q$.

We define a minimax robust filter $f_\mathrm{r}$ as the filter that solves
\begin{eqnarray*}
\max_{f \in F} \inf_{q \in Q} U(f, q).
\end{eqnarray*}
Its dual is a least favorable operating point $q_\mathrm{l}$ defined as the operating point that solves
\begin{eqnarray*}
\min_{q \in Q} \sup_{f \in F} U(f, q).
\end{eqnarray*}

A saddle point solution to the game $(F, Q, U)$ is $(f_\mathrm{l}, q_\mathrm{l}) \in F \times Q$ that satisfies
\begin{eqnarray}
U(f, q_\mathrm{l}) \leq U(f_\mathrm{l}, q_\mathrm{l}) \leq U(f_\mathrm{l}, q),
\end{eqnarray}
for every $(f, q) \in F \times Q$. Note that if a game has a saddle point, then the corresponding filter is the robust minimax filter for the game, and furthermore the corresponding operating point is the least favorable operating point.

A pair $(f_\mathrm{l}, q_\mathrm{l}) \in F \times Q$ is called a regular pair if, for every $q \in Q$ such that $q_\alpha := (1 - \alpha)q_\mathrm{l} + \alpha q \in Q$ for $\alpha \in [0, 1]$, we have
\begin{eqnarray}
\sup_{f \in F} U(f, q_\alpha) - U(f_\mathrm{l}, q_\alpha) = o(\alpha),
\end{eqnarray}
where $o(\alpha)/\alpha \rightarrow 0$ as $\alpha \rightarrow 0$.

The following theorem is the key result that we utilize in Section \ref{sec:capacity}.
\begin{thm}\cite[Theorem 2.1]{verdu84:it}
\label{thm:verdu84}
Suppose that the game $(F, Q, U)$ is such that
\begin{itemize}
\item[(a)] $Q$ is a convex set,
\item[(b)] $U(f, \cdot)$ is convex on $Q$ for every $f \in F$.
\end{itemize}
Then, if $(f_\mathrm{l}, q_\mathrm{l})$ is a regular pair for $(F, Q, U)$, the following are equivalent:
\begin{itemize}
\item[(1)] $q_\mathrm{l}$ is a least favorable operating point for $(F, Q, U)$,
\item[(2)] $(f_\mathrm{l}, q_\mathrm{l})$ is a saddle point solution for $(F, Q, U)$.
\end{itemize}
\end{thm}

In contrast to typical minimax theorems (see, {\it e.g.}, \cite{simons95:survey}), Theorem \ref{thm:verdu84} lifts the (quasi-)concavity constraint of $U(f, q)$ in $f$, and its validity in fact requires neither certain topological properties ({\it e.g.}, compactness) of $\mathcal{F}$ and $\mathcal{Q}$, nor (semi-)continuity properties of $U(f, q)$ on $F$ and $Q$. Indeed, the only essential requirement is the existence of the least favorable operating point. Therefore under certain circumstances, Theorem \ref{thm:verdu84} appears quite convenient to utilize.

A systematic procedure can be followed in order to apply Theorem \ref{thm:verdu84} for a game $(F, Q, U)$:
\begin{enumerate}
\item Verify that the assumptions in Theorem \ref{thm:verdu84} are satisfied.
\item Find the least favorable operating point $q_\mathrm{l}$ and the corresponding optimal filter $f^\ast(q_\mathrm{l})$ that solves $\sup_{f \in F} U(f, q_\mathrm{l})$.
\item Verify that the solved $\left(f^\ast(q_\mathrm{l}), q_\mathrm{l}\right)$ is a regular pair. The saddle point solution $\left(f^\ast(q_\mathrm{l}), q_\mathrm{l}\right)$ then yields the minimax robust filter $f_\mathrm{r} = f^\ast(q_\mathrm{l})$.
\end{enumerate}

\subsection{Proof of the equivalence between (\ref{eqn:rate-raw}) and (\ref{eqn:rate-1})}
\label{app:interchange}

We follow Appendix \ref{app:general} to prove the equivalence between (\ref{eqn:rate-raw}) and (\ref{eqn:rate-1}). In the problem, $g(\cdot)$ is the filter, $(P_{s, \mathrm{f}}, P_{s, \mathrm{c}})$ is the operating point, and
\begin{eqnarray}
\label{eqn:payoff-def}
U\left(g, (P_{s, \mathrm{f}}, P_{s, \mathrm{c}})\right) := \sup_{\theta \geq 0} \{ \theta \mathbf{E}_{P_{s, \mathrm{c}}} [g(\rvY)] - \log \mathbf{E}_{P_{s, \mathrm{f}}} \left\{\exp[\theta g(\rvY)]\right\} \}
\end{eqnarray}
is the payoff function.

Let us first verify the required conditions for the game. By assumption, the set of allowable operating points, $\mathcal{P}_{\svx_\mathrm{f}} \times \mathcal{P}_{\svx_\mathrm{c}}$, is convex. We need to show that for every $g(\cdot)$, the payoff function is convex on $\mathcal{P}_{\svx_\mathrm{f}} \times \mathcal{P}_{\svx_\mathrm{c}}$. Consider two arbitrary $(P_{s, \mathrm{f}}, P_{s, \mathrm{c}})$ and $(P^\prime_{s, \mathrm{f}}, P^\prime_{s, \mathrm{c}})$ in $\mathcal{P}_{\svx_\mathrm{f}} \times \mathcal{P}_{\svx_\mathrm{c}}$, and for every $\alpha \in [0, 1]$ the convex combination $(P_{s, \mathrm{f}, \alpha}, P_{s, \mathrm{c}, \alpha}) := \left(\alpha P_{s, \mathrm{f}} + (1 - \alpha) P^\prime_{s, \mathrm{f}}, \alpha P_{s, \mathrm{c}} + (1 - \alpha) P^\prime_{s, \mathrm{c}}\right)$. We have
\begin{eqnarray}
U\left(g, (P_{s, \mathrm{f}, \alpha}, P_{s, \mathrm{c}, \alpha})\right) &=& \sup_{\theta \geq 0} \{\theta \mathbf{E}_{P_{s, \mathrm{c}, \alpha}} [g(\rvY)] - \log \mathbf{E}_{P_{s, \mathrm{f}, \alpha}} \left\{\exp[\theta g(\rvY)]\right\}\}\nonumber\\
&=& \sup_{\theta\geq 0} \left\{ \alpha \theta \mathbf{E}_{P_{s, \mathrm{c}}} [g(\rvY)] + (1 - \alpha) \theta \mathbf{E}_{P^\prime_{s, \mathrm{c}}} [g(\rvY)] \right.\nonumber\\&& \left.- \log \left\{
\alpha \mathbf{E}_{P_{s, \mathrm{f}}} \left\{\exp[\theta g(\rvY)]\right\} + (1 - \alpha) \mathbf{E}_{P^\prime_{s, \mathrm{f}}} \left\{\exp[\theta g(\rvY)]\right\}
\right\}\right\}\nonumber\\
&\leq& \sup_{\theta\geq 0}\left\{ \alpha \theta \mathbf{E}_{P_{s, \mathrm{c}}} [g(\rvY)] + (1 - \alpha) \theta \mathbf{E}_{P^\prime_{s, \mathrm{c}}} [g(\rvY)]\right. \nonumber\\&&\left. - \alpha \log\mathbf{E}_{P_{s, \mathrm{f}}} \left\{\exp[\theta g(\rvY)]\right\} - (1 - \alpha) \log\mathbf{E}_{P^\prime_{s, \mathrm{f}}} \left\{\exp[\theta g(\rvY)]\right\}\right\}\nonumber\\
&\leq& \alpha U\left(g, (P_{s, \mathrm{f}}, P_{s, \mathrm{c}})\right) + (1 - \alpha) U\left(g, (P^\prime_{s, \mathrm{f}}, P^\prime_{s, \mathrm{c}})\right),
\end{eqnarray}
where the first inequality follows from the concavity property of logarithmic functions, and the second inequality follows from the supremum operation.

The next step is to find the least favorable operating point and the corresponding optimal filter. This is solved in Appendix \ref{app:optimal}. The least favorable operating point is the distributions pair $(P_\mathrm{f}^\ast, P_\mathrm{c}^\ast)$ that minimizes
\begin{eqnarray*}
{D}\left(P_{s, \mathrm{c}}\| P_{s, \mathrm{f}}\right), \quad s \in \mathcal{S},
\end{eqnarray*}
and the corresponding optimal filter is
\begin{eqnarray*}
g^\ast(\svy) = \log\frac{P_\mathrm{c}^\ast(\svy)}{P_\mathrm{f}^\ast(\svy)}, \quad \mathrm{and}\; \theta^\ast = 1.
\end{eqnarray*}

Finally, we need to verify that $\left(g^\ast, (P_\mathrm{f}^\ast, P_\mathrm{c}^\ast)\right)$ is a regular pair. Consider for every $(Q_\mathrm{f}, Q_\mathrm{c}) \in \mathcal{P}_{\svx_\mathrm{f}} \times \mathcal{P}_{\svx_\mathrm{c}}$ the neighboring operating point
\begin{eqnarray*}
\left(P_{\mathrm{f}, \alpha}, P_{\mathrm{c}, \alpha}\right) = \left(
(1 - \alpha)P^\ast_\mathrm{f} + \alpha Q_\mathrm{f}, (1 - \alpha)P^\ast_\mathrm{c} + \alpha Q_\mathrm{c}
\right),
\end{eqnarray*}
for small $\alpha \in [0, 1]$.

On one hand, by Appendix \ref{app:optimal},
\begin{eqnarray}
\label{eqn:equivalence-1}
\sup_{(\theta, g)} U\left(g, \left(P_{\mathrm{f}, \alpha}, P_{\mathrm{c}, \alpha}\right)\right) = {D}\left(
P_{\mathrm{c}, \alpha} \| P_{\mathrm{f}, \alpha}
\right) = \mathbf{E}_{P_{\mathrm{c}, \alpha}} \left[
\log \frac{P_{\mathrm{c}, \alpha}(\rvY)}{P_{\mathrm{f}, \alpha}(\rvY)}
\right].
\end{eqnarray}
Expanding the right hand side of (\ref{eqn:equivalence-1}) at $\alpha = 0$, we have
\begin{eqnarray}
\left.\mathbf{E}_{P_{\mathrm{c}, \alpha}} \left[
\log \frac{P_{\mathrm{c}, \alpha}(\rvY)}{P_{\mathrm{f}, \alpha}(\rvY)}
\right]\right|_{\alpha = 0} &=& {D}(P_\mathrm{c}^\ast\|P_\mathrm{f}^\ast);\\
\frac{d}{d\alpha}\mathbf{E}_{P_{\mathrm{c}, \alpha}} \left[
\log \frac{P_{\mathrm{c}, \alpha}(\rvY)}{P_{\mathrm{f}, \alpha}(\rvY)}
\right] &=& \mathbf{E}_{Q_\mathrm{c}} \left[
\log \frac{P_{\mathrm{c}, \alpha}(\rvY)}{P_{\mathrm{f}, \alpha}(\rvY)}
\right] - \mathbf{E}_{P^\ast_\mathrm{c}} \left[
\log \frac{P_{\mathrm{c}, \alpha}(\rvY)}{P_{\mathrm{f}, \alpha}(\rvY)}
\right]\nonumber\\
&& + \mathbf{E}_{P_{\mathrm{c}, \alpha}} \left[
\frac{Q_\mathrm{c}(\rvy) - P^\ast_\mathrm{c}(\rvy)}{P_{\mathrm{c}, \alpha}(\rvy)} - \frac{Q_\mathrm{f}(\rvy) - P^\ast_\mathrm{f}(\rvy)}{P_{\mathrm{f}, \alpha}(\rvy)}
\right]\\
\left.\frac{d}{d\alpha}\mathbf{E}_{P_{\mathrm{c}, \alpha}} \left[
\log \frac{P_{\mathrm{c}, \alpha}(\rvY)}{P_{\mathrm{f}, \alpha}(\rvY)}
\right]\right|_{\alpha = 0} &=& \mathbf{E}_{Q_\mathrm{c}} \left[
\log \frac{P_\mathrm{c}^\ast(\rvY)}{P_\mathrm{f}^\ast(\rvY)}
\right] - {D}(P_\mathrm{c}^\ast\|P_\mathrm{f}^\ast) + 1 - \mathbf{E}_{Q_\mathrm{f}}\left[\frac{P_\mathrm{c}^\ast(\rvY)}{P_\mathrm{f}^\ast(\rvY)}\right].
\end{eqnarray}
Hence the first-order expansion of (\ref{eqn:equivalence-1}) is
\begin{eqnarray}
\label{eqn:regular-1}
&&\sup_{(\theta, g)} U\left(g, \left(P_{\mathrm{f}, \alpha}, P_{\mathrm{c}, \alpha}\right)\right) \nonumber\\&=& {D}(P_\mathrm{c}^\ast\|P_\mathrm{f}^\ast)
 + \left\{
 \mathbf{E}_{Q_\mathrm{c}} \left[
\log \frac{P_\mathrm{c}^\ast(\rvY)}{P_\mathrm{f}^\ast(\rvY)}
\right] - {D}(P_\mathrm{c}^\ast\|P_\mathrm{f}^\ast) + 1 - \mathbf{E}_{Q_\mathrm{f}}\left[\frac{P_\mathrm{c}^\ast(\rvY)}{P_\mathrm{f}^\ast(\rvY)}\right]
 \right\} \cdot \alpha + o(\alpha).
\end{eqnarray}

On the other hand, noting that for the optimal filter $g^\ast$ we have $\theta^\ast = 1$, hence
\begin{eqnarray}
\label{eqn:regular-2}
&&U\left(g^\ast, \left(P_{\mathrm{f}, \alpha}, P_{\mathrm{c}, \alpha}\right)\right) \nonumber\\&=& \mathbf{E}_{P_{\mathrm{c}, \alpha}} \left[
\log\frac{P_\mathrm{c}^\ast(\rvY)}{P_\mathrm{f}^\ast(\rvY)}
\right] - \log\mathbf{E}_{P_{\mathrm{f}, \alpha}}\left[\frac{P_\mathrm{c}^\ast(\rvY)}{P_\mathrm{f}^\ast(\rvY)}\right]\nonumber\\
&=& (1 - \alpha) {D}(P_\mathrm{c}^\ast\|P_\mathrm{f}^\ast) + \alpha \mathbf{E}_{Q_\mathrm{c}} \left[
\log \frac{P_\mathrm{c}^\ast(\rvY)}{P_\mathrm{f}^\ast(\rvY)}
\right] - \log\left\{
1 - \alpha + \alpha \mathbf{E}_{Q_\mathrm{f}}\left[\frac{P_\mathrm{c}^\ast(\rvY)}{P_\mathrm{f}^\ast(\rvY)}\right]
\right\}\nonumber\\
&=& (1 - \alpha) {D}(P_\mathrm{c}^\ast\|P_\mathrm{f}^\ast) + \alpha \mathbf{E}_{Q_\mathrm{c}} \left[
\log \frac{P_\mathrm{c}^\ast(\rvY)}{P_\mathrm{f}^\ast(\rvY)}
\right] - \left\{
\mathbf{E}_{Q_\mathrm{f}}\left[\frac{P_\mathrm{c}^\ast(\rvY)}{P_\mathrm{f}^\ast(\rvY)}\right] - 1
\right\} \cdot \alpha + o(\alpha),
\end{eqnarray}
where we have used $\log(1 + t) = t + o(t)$ for $|t| \approx 0$.

Comparing (\ref{eqn:regular-1}) and (\ref{eqn:regular-2}), we find that all the terms, except $o(\alpha)$, cancel out. So that $\left(g^\ast, (P_\mathrm{f}^\ast, P_\mathrm{c}^\ast)\right)$ is a regular pair, and by the saddle point property we establish the validity of (\ref{eqn:rate-1}).

\subsection{Proof of (\ref{eqn:sup-g}) and (\ref{eqn:optimum-g})}
\label{app:optimal}

In order to maximize $\left\{\theta \mathbf{E}_{P_{s, \mathrm{c}}}[g(\rvY)] - \log \mathbf{E}_{P_{s, \mathrm{f}}}\{\exp[\theta g(\rvY)]\}\right\}$ with respect to $\theta$ and $g(\cdot)$, we let its variation regarding $g(\cdot)$ be zero, {\it i.e.},
\begin{eqnarray}
\theta P_{s, \mathrm{c}}(\svy) - \frac{\theta P_{s, \mathrm{f}}(\svy)\exp[\theta g(\svy)]}{\mathbf{E}_{P_{s, \mathrm{f}}}\{\exp[\theta g(\rvY)]\}} = 0, \quad \mathrm{for\;all\;} \svy \in \mathcal{Y}.
\end{eqnarray}
That is,
\begin{eqnarray}
\label{eqn:optimal-gy}
g(\svy) = \frac{1}{\theta}\log\frac{P_{s, \mathrm{c}}(\svy)}{P_{s, \mathrm{f}}(\svy)} + \frac{1}{\theta}\log \mathbf{E}_{P_{s, \mathrm{f}}}\{\exp[\theta g(\rvY)]\}.
\end{eqnarray}
It can then be verified that a solution to (\ref{eqn:optimal-gy}) is
\begin{eqnarray}
g(\svy) = \log\frac{P_{s, \mathrm{c}}(\svy)}{P_{s, \mathrm{f}}(\svy)},\quad\mathrm{and}\;\theta = 1,
\end{eqnarray}
which then yields the maximum value ${D}(P_{s, \mathrm{c}}\|P_{s, \mathrm{f}})$. The result is intuitively apparent, because the log-likelihood ratio is a sufficient statistic conditioned upon a given channel state realization.

\section*{Acknowledgment}
The authors wish to thank the anonymous reviewers for their detailed comments on this work, and an anonymous reviewer of an earlier version of this work for suggesting (\ref{eqn:cpuc-lp}) in Proposition \ref{prop:cpuc}. The first author (W.Z.) also wishes to thank J. Nicholas Laneman for encouragement and suggestions throughout the preparation of this paper.

\bibliographystyle{ieee}
\bibliography{../it07133robust}

\end{document}